\documentclass[12pt]{elsarticle}

\usepackage{graphicx}
\usepackage{amsmath}
\usepackage{amsthm}
\usepackage{amsfonts}
\usepackage[brazil]{babel}
\usepackage[utf8]{inputenc}

\newcounter{bla}

\journal{Computer Physics Communications}

\newtheorem{teor}{Theorem}[section]
\newtheorem{cor}{Corollary}[section]
\newtheorem{obs}{Remark}[section]
\newtheorem{defin}{Definition}[section]

\newtheorem{algor}{Algorithm}[section]

\begin{document}

\title{\uppercase{Dealing with Rational Second Order Ordinary Differential Equations where both Darboux and Lie Find It Difficult: The $S$-function Method}}

\author[FAETEC]{J. Avellar}
\ead{mscardoso@gmail.com.br}

\author[uerj]{M.S. Cardoso}
\ead{mscardoso@gmail.com.br}

\author[uerj]{L.G.S. Duarte}
\ead{lgsduarte@gmail.com.br}

\author[uerj]{L.A.C.P. da Mota\corref{cor1}}
\ead{lacpdamota@uerj.br or lacpdamota@dft.if.uerj.br}

\cortext[cor1]{Corresponding author}

\address[uerj]{Universidade do Estado do Rio de Janeiro,
{\it Instituto de F\'{\i}sica, Depto. de F\'{\i}sica Te\'orica},
{\it 20559-900 Rio de Janeiro -- RJ, Brazil}}

\address[FAETEC]{ Funda\c c\~ao de Apoio \`a Escola T\'ecnica,
{\it  RJ, Brazil}}

\begin{abstract}
Here we present a new approach to search for first order invariants (first integrals) of rational second order ordinary differential equations. This method is an alternative to the Darbouxian and symmetry approaches. Our procedure can succeed in many cases where  these two approaches fail. We also present here a Maple implementation of the theoretical results and methods, hereby introduced, in a computational package -- {\it InSyDE}. The package is designed, apart from materializing the algorithms presented, to provide a set of tools to allow the user to analyse the intermediary steps of the process.
\end{abstract}


\maketitle

\newpage
\bigskip
\hspace{1pc}
{\bf PROGRAM SUMMARY}
\bigskip

\begin{footnotesize}
\noindent
{\em Title of the software package:}  {\em InSyDE} -- Invariants and Symmetries of (racional second order ordinary) Differential Equations.   \\[10pt]
{\em Catalogue number:} (supplied by Elsevier)                \\[10pt]
{\em Software obtainable from:} CPC Program Library, Queen's
University of Belfast, N. Ireland.
\\[10pt]
{\em Licensing provisions:} none  \\[10pt]
{\em Operating systems under which the program has been tested:}
Windows 8.
\\[10pt]
{\em Programming languages used:} Maple 17.
\\[10pt]
{\em Memory required to execute with typical data:}  200 Megabytes. \\[10pt]
{\em No. of lines in distributed program, including On-Line Help,
etc.:} 537   \\[10pt]
{\em Nature of mathematical problem}\\
Search for first integrals of rational 2ODEs.
\\[10pt]
{\em Methods of solution}\\
The method of solution is based on an algorithm described in this paper.
\\[10pt]
{\em Restrictions concerning the complexity of the problem}\\
If the rational 2ODE that is being analysed presents a very high degree in ($x,y,z$), then the method may not work well.
\\[10pt]
{\em Typical running time}\\
This depends strongly on the 2ODE that is being analysed.
\\[10pt]
{\em Unusual features of the program}\\
Our implementation can find first integrals in many cases where the rational 2ODE under study can not be reduced by other powerful solvers. Besides that, the package presents some useful research commands.
\end{footnotesize}
\newpage
\hspace{1pc}
{\bf LONG WRITE-UP}

Since the advent of the Newtonian approach to mechanics, the differential equations (DEs) became the main framework to model most phenomena. In the beginning, the way to deal with the solving of DEs was {\em classificatory}, i.e., if someone discovered a new method to solve a particular type of differential equation, it would add this method to a long list of methods already cataloged. This way of doing things lasted until the late nineteenth century when the appearance of Lie and Darboux works \cite{Lie,Dar} inaugurated a more general way of looking at the problem. The Lie theory is centered on the concept of symmetry: if a DE remains invariant (it does not change its shape) when subjected to a continuous group of transformations\footnote{These groups are known today as {\em Lie groups}.} (these groups are called symmetry groups or just symmetries of the DE) then we can reduce its order \cite{BluAnc,Ibr,Olv,Sch,Ste}. The method of Darboux, on the other hand, was based on the concept of {\em invariant algebraic curve}. The polynomials that define these invariant algebraic curves (Darboux polynomials) are eigenfunctions of a differential operator associated with the DE or with the system of DEs. If such a system has a certain number of invariant algebraic curves, then it presents an algebraic first integral.

These two theories are generalists in the sense that they do not target DEs of a type defined by its solution method: the Lie method can be applied to ordinary differential equations (ODEs) of any order, to systems of ODEs, to partial differential equations (PDEs), to PDEs systems etc; the Darbouxian approaches dealt with polynomial systems of 1ODEs presenting algebraic first integrals.
However, despite being very rich and widely applicable theories, Lie and Darboux methods also possessed some drawbacks: the Lie method did not provide an algorithm to calculate the symmetries\footnote{Especially when ODEs do not present point symmetries. In this case we can not separate the determining equation (the PDE that the symmetries should obey) in the derivatives of the dependent variable to obtain an overdetermined system of PDEs.} and, sometimes, we can face a PDE for the symmetries that is much more complicated to solve than the ODE in question. On the other hand, Darbouxian methods were (at first) applied only to polynomial systems of 1ODEs.

To overcome these points several approaches have been developed:
\begin{itemize}
\item For dealing with ODEs using symmetry methods L.G.S. Duarte and L.A.C.P. da Mota \cite{Noscpc1997,Noscpc1998} created a series of heuristics to find symmetries of first and second order ODEs; P. J. Olver introduced the concept of exponencial vector field (see \cite{Olv}, p. 185); B. Abraham-Shrauner, A. Guo, K.S. Govinder, P.G.L. Leach, F.M. Mahomed, A.A. Adam, M. L. Gandarias, M. S. Bruzón, M. Senthilvelan and others worked with the concept of hiden and non local symetries \cite{AbrGuo,AbrGuo2,AbrGovLea,Abr,GovLea,AdaMah,GanBru,GanBruSen}. C. Muriel and J.L. Romero have developed the concept of $\lambda$-symmetry \cite{MurRom,MurRom2} that inaugurates a rich research area (see \cite{MurRom3,MurRom4,CicGaeWal,CicGaeMor} and references therein). E. Pucci and G. Saccomandi created the concept of telescopical symmetry \cite{PucSac}. Another great approach was brought by M.C. Nucci that used the concept of Jacobi last multiplier (see \cite{Nuc,Nuc2}) in a very clever way.
\item In the Darbouxian context, Cairó and Llibre \cite{CaiLli} pointed out that by using the exponential factors, introduced by Cristopher \cite{Chr}, the Darboux methods could be generalized to deal with elementary (rather than just algebraic) first integrals. In 1983, Prelle and Singer \cite{PreSin} found a semialgorithmic approach to find elementary first integrals of 2D vector fields (or, equivalently, rational 1ODEs). Because of its remarkable characteristics the Darbouxian approach has generated many extensions \cite{Sht,Col,Sin,Chr2,ChrLli,Lli,Nosjpa2002-1,Nosjpa2002-2,Noscpc2002,Nosjcam2005,Nosjpa2001,Noscpc2007,Nosamc2007,Nosjmp2009,Nosjpa2010,LliZha}. In particular, in \cite{Chr2,ChrLli,Nosjpa2002-1,Nosjpa2002-2,Nosjcam2005,Noscpc2007} the method was extended to deal with rational 1ODEs presenting Liouvillian solutions. In \cite{Nosjpa2001,Nosamc2007,Nosjmp2009}, it was extended to deal with rational 2ODEs that present at least one elementary first integral. In \cite{Nosjpa2010,LliZha} it deals with polynomial systems of 1ODEs in more than two independent variables.
\end{itemize}

In this work we have developed a method (to deal with rational 2ODEs) that could be viewed as a mix between symmetry methods and the Darbouxian approaches. The main motivation for the development of this method is that some rational 2ODEs have integrating factors formed by Darboux polynomials of relatively high degree (which means, in practice we can not determine them with the memory of a standard personal computer) and, on the other hand, do not present point symmetries (which means that, in practice, it could be very hard to determine a dynamical or non local symmetry).

\section{The $S$-functions and the associated 1ODEs of a 2ODE}
\label{SandS}
\hspace\parindent
In \cite{Nosjpa2001} we developed the concept of $S$-function associated with a 2ODE through an invariant (a first integral). The basic idea was, in short, to `complete' the differential $dI$ of the first integral $I$ in the basis $\{dx,dy,dz\}$ (where $z \equiv y'$), in order to construct a generalization of the Prelle-Singer procedure \cite{PreSin}.
In this section we show how to use the $S$-function to deal with rational 2ODEs that are difficult to solve/reduce by using the Darboux or Lie approaches:
\begin{itemize}
\item First we present the concept of $S$-function associated with a 2ODE. The $S$-function is linked with the fact that we can express the differential of a first integral (of a 2ODE) as a linear combination of two 1-forms that are zero over the solutions of the 2ODE.
\item Next, we present the concept of associated 1ODE (introduced in \cite{Nosjmp2009}) and show how its solution is related to the solving/reducing of rational 2ODEs presenting Liouvillian first integrals.
\end{itemize}

\subsection{The $S$-function}
\label{Sfunc}
\hspace\parindent
Consider the rational 2ODE given by
\begin{equation}
\label{2oder1}
z'=\frac{dz}{dx}=\phi(x,y,z)=\frac{M(x,y,z)}{N(x,y,z)},  \,\,(z \equiv y'),
\end{equation}
where $M$ and $N$ are coprime polynomials in $(x,y,z)$.

\begin{defin}
A function $I(x,y,z)$ is called {\bf first integral} of the 2ODE {\em (\ref{2oder1})} if $I$ is constant over the solutions of {\em (\ref{2oder1})}.
\end{defin}

\begin{obs} If $I(x,y,z)$ is a first integral of the 2ODE {\em (\ref{2oder1})} then, over the solution curves of {\em (\ref{2oder1})}, the exact 1-form $\omega:=dI=I_x\,dx+I_y\,dy+I_{z}\,dz\,$ is null.
\end{obs}
\noindent
Over the solution curves of (\ref{2oder1}), we have two independent null 1-forms:
\begin{eqnarray}
\label{alfa2}
\alpha & := & \phi\,dx-dz, \\
\label{beta2}
\beta & := & z\,dx-dy.
\end{eqnarray}
So, the 1-form $\omega$ is in the vector space sppaned by the 1-forms $\alpha$ and $\beta$, i.e., we can write $\omega = r_1 (x,y,z)\,\alpha+r_2 (x,y,z)\,\beta$:
\begin{eqnarray}
\label{eqform2edo1}
dI &=& I_x\,dx+I_y\,dy+I_{z}\,dy = r_1\,(\phi\,dx - dz)\,+r_2\,(z\,dx - dy) \nonumber \\
    &=& (r_1\,\phi\,+r_2\,z)dx + (-r_2)dy + (-r_1)dz.
\end{eqnarray}
Therefore, we have that\footnote{from the knowledge of $r_1$ and $r_2$ satisfying (\ref{Is2edo}), we can construct a first integral via quadratures:
\begin{eqnarray}
I(x,y,z) & = & \int I_x\,dx+\int\!\left[ I_y - \partial{}{y}\int I_x\,dx\right]\,dy + \nonumber \\
&&\int\!\left( I_{z} - \partial{}{z}\left[ \int I_x\,dx+\int\!\left[ I_y - \partial{}{y}\int I_x\,dx\right]\,dy   \right]
\right) \,dz \nonumber \\ \Rightarrow \nonumber \\
I(x,y,z) & = & \int \!\left(r_1\phi+r_2\,z\right)\,dx-\int\!\left[r_2+\partial{}{y}\int \!\left(r_1\phi+r_2\,z\right)\,dx\right]\,dy- \nonumber
\end{eqnarray}
\begin{equation}
\label{inv2edo}
\int\!\left\{r_1+\partial{}{z}\left[\int\!\left(r_1\phi+r_2\,z\right)\,dx-\int\!\left[r_2+\partial{}{y}\int\!
\left(r_1\phi+r_2\,z\right)\,dx\right]dy\right]\right\}dz.
\end{equation}}:
\begin{eqnarray}
\label{Is2edo}
I_x & = & r_1\,\phi+r_2\,z, \nonumber \\
I_y & = & -r_2,  \\
I_z & = & -r_1. \nonumber
\end{eqnarray}

\noindent
If $I(x,y,z)$ is a first integral of the 2ODE (\ref{2oder1}) we can write
\begin{eqnarray}
\label{eqform2edo2}
dI &=& (r_1\,\phi\,+r_2\,z)dx + (-r_2)dy + (-r_1)dz = \nonumber \\
    &=& r_1\,\left[(\phi\,+\frac{r_2}{r_1}\,z)dx - \frac{r_2}{r_1}dy - dz\right].
\end{eqnarray}

\noindent
If we call $\,r_1 \equiv R\,$ and $\,\frac{r_2}{r_1} \equiv S\,$, we can write
\begin{equation}
\label{dIrands}
dI = R\,\left[(\phi\,+z\,S)dx - S\,dy - dz\right].
\end{equation}

\begin{defin}
\label{fatint1form}
Let $\gamma$ be a 1-form. We say that $R$ is an {\bf integrating factor} for the 1-form $\gamma$ if $R\,\gamma$ is an exact 1-form.
\end{defin}

\begin{defin}
Let $I$ be a first integral of the {\em 2ODE (\ref{2oder1})}. The function defined by $S := I_y/I_z$ is called a \mbox{\boldmath $S$}{\bf -function} associated with the {\em 2ODE} through the first integral $I$.
\end{defin}

\begin{obs}
From the above definitions and in view of {\em (\ref{dIrands})} we can see that $R$ is an integrating factor for the 1-form $\,(\phi\,+z\,S)dx - S\,dy - dz\,$ and $S$ is a $S$-function associated with the {\em 2ODE (\ref{2oder1})} through $I$ .
\end{obs}
\noindent
So, we can write (\ref{Is2edo}) as
\begin{eqnarray}
\label{Is2edo2}
I_x & = & R\,(\phi+z\,S), \nonumber \\
I_y & = & -R\,S,  \\
I_z & = & -R. \nonumber
\end{eqnarray}
We can (with some basic algebra) write the compatibility conditions of (\ref{Is2edo2}) in a very handy format. This is expressed in the following result:

\begin{teor}
\label{concom}
Let $\,I(x,y,z)\,$ be a first integral of the {\em 2ODE (\ref{2oder1})}. If $S$ is the $S$-function associated to the {\em 2ODE (\ref{2oder1})} through $I$, then we can write the compatibility conditions for {\em (\ref{Is2edo2})} as
\begin{eqnarray}
\label{eqproof1r}
&& D_x[R] +R\,(S+\phi_z)=0\,, \\
\label{eqproof2r}
&& S\,D_x[R] + R\,( D_x[S]+\phi_y)=0\,, \\
\label{eqproof3r}
&& -(R_z\,S+R\,S_z) + R_y=0\,,
\end{eqnarray}
where $\,D_x := \partial_x + z\,\partial_y + \phi(x,y,z)\,\partial_z\,$.
\end{teor}

\noindent
{\bf Proof of Theorem \ref{concom}:} From the compatibility conditions $(I_{xy}-I_{yx}=0, I_{xz}-I_{zx}=0$ and
$I_{yz}-I_{zy}=0 )$, we have:
\begin{eqnarray}
\label{condcomp1}
&& R_y\,(\phi+z\,S)+R\,(\phi_y+S_y\,z) + (S_x\,R + S\,R_x)=0\,,\\
\label{condcomp2}
&&R_z\,(\phi+z\,S)+R\,(\phi_z+S_z\,z+S) + R_x=0\,, \\
\label{condcomp3}
&&- ( R_z\,S+R\,S_z) + R_y=0\,.
\end{eqnarray}

\noindent
The condition (\ref{eqproof3r}) is (already) expressed by eq.(\ref{condcomp3}).
Eq.(\ref{condcomp2}) plus eq.(\ref{condcomp3}) times $z$ results
\begin{equation}
\label{eqproof1}
R_x + z\,R_y+\phi\,R_{z}+R\,(\phi_{z}+S) =0\,,
\end{equation}

\noindent
and eq.(\ref{condcomp1}) minus eq.(\ref{condcomp3}) times $\phi$ results
\begin{equation}
\label{eqproof2}
S\,(R_x + z\,R_y+\phi\,R_{z})+R\,(S_x + z\,S_y+\phi\,S_{z})+R\,\phi_y =0\,.
\end{equation}

\noindent
Equations (\ref{eqproof1}) and (\ref{eqproof2}) can be written, respectively, as
\begin{eqnarray}
\label{eqproof1f}
&&D_x[R]+R\,(\phi_{z}+S) =0\,,\\
\label{eqproof2f}
&&S\,D_x[R]+R\,D_x[S]+R\,\phi_y =0\,. \,\,\, \Box
\end{eqnarray}

\begin{cor}
\label{corS}
Let $S$ be a $S$-function associated with the {\em 2ODE} $\,z'=\phi(x,y,z)$. Then $S$ obeys the following equation:
\begin{equation}
\label{eqS}
D_x[S]=S^2+\phi_{z}\,S-\phi_y\,.
\end{equation}
\end{cor}

\noindent
{\bf Proof of Corollary \ref{corS}:} Isolating $\,D_x[R]/R\,$ in (\ref{eqproof1r}) e substituting in (\ref{eqproof2r}) we obtain (\ref{eqS}).  $\Box$

\bigskip

\subsection{The Associated 1ODEs}
\label{conectSandS}
\hspace\parindent
From (\ref{eqS}) we can see that a $S$-function associated with the rational 2ODE (\ref{2oder1}) satisfies a quasilinear 1PDE in the variables $(x,y,z)$:
\begin{equation}
\label{PDES}
D_x[S]=S_x + z\,S_y + \phi(x,y,z)\,S_{z}=S^2+\phi_z\,S-\phi_y\,.
\end{equation}
Over the solutions of the 2ODE (\ref{2oder1}) we have that $y=y(x)$ and $z=z(x)$ and, therefore, the operator $D_x$ is, formally, ${\frac{d}{dx}}$. So, formally, over the solutions of the 2ODE (\ref{2oder1}) we can write the 1PDE (\ref{PDES}) as a Riccati 1ODE:
\begin{equation}
\label{1edoS}
\frac{ds}{dx}=s^2+\phi_z\,s-\phi_y.
\end{equation}
It is of common knowledge that the transformation
\begin{equation}
\label{ricctol2ode}
y(x)=-\frac{r'(x)}{f(x)\,r(x)}
\end{equation}
changes the Riccati equation $\,y'(x)=f(x)\,y(x)^2+g(x)\,y(x)+h(x)\,$ into the linear 2ODE
\begin{equation}
\label{l2odefricc}
r''=\frac{f'(x)+g(x)\,f(x)}{f(x)}\,r'-f(x)\,h(x)\,r.
\end{equation}
So, with the transformation
\begin{equation}
\label{ricctol2ode2}
s(x)=-{\frac{\frac{d}{dx}w(x)}{w(x)}},
\end{equation}
the Riccati 1ODE (\ref{1edoS}) turns (over the solutions of the 2ODE (\ref{2oder1})) into the following homogeneous linear 2ODE:
\begin{equation}
\frac{d^2 w}{dx^2}= {\phi_z}\,\frac{dw}{dx} + \phi_y\,w.
\end{equation}
We can use the formal equivalence $D_x\,\sim\,{\frac{d}{dx}}$ to produce a connection between the $S$-functions and the symmetries (written in a particular form) of the 2ODE: let's make the transformation (the formal analogous of the transformation (\ref{ricctol2ode}))
\begin{equation}
\label{stonu}
S=-\frac{D_x[\nu]}{\nu}
\end{equation}
into equation (\ref{PDES}). We obtain:
\begin{equation}
\label{2PDEnu}
D_x^2[\nu]=\phi_z\,D_x[\nu]+\phi_y\,\nu.
\end{equation}
The equation (\ref{2PDEnu}) is the symmetry condition for $\,\nu\,$ to be the infinitesimal that defines a symmetry generator in the evolutionary form. So, we can enunciate the following result:
\begin{teor}
\label{connectSSy}
Let $\nu$ be a function of $(x,y,z)$ such that $[0,\nu]$ defines a symmetry of the {\em 2ODE (\ref{2oder1})} in the evolutionary form, i.e., $X_e := \nu\,\partial_y\,$ generates a symmetry transformation for  {\em (\ref{2oder1})}. Then the function defined by $S=-D_x[\nu]/\nu$ is a $S$-function associated with the {\em 2ODE (\ref{2oder1})}.
\end{teor}
\noindent
{\bf Proof of Theorem \ref{connectSSy}:} The first extension of $X_e$ is $\,X^{(1)}_e = \nu\,\partial_y+D_x[\nu]\,\partial_z$. Let $I$ be a first integral of the 2ODE (\ref{2oder1}), such that (without loss of generality) ${X_e}^{(1)}[I]=0$. So, $\nu\,I_y+D_x[\nu]\,I_{z}=0 \,\Rightarrow\, I_y/I_z=-D_x[\nu]/\nu. \,\,\,\Box$

\begin{cor}
\label{connectSyS}
Let $S$ be a $S$-function associated with the {\em 2ODE (\ref{2oder1})}. Then, the function $\nu$ given by
\begin{equation}
\label{nuS}
\nu \equiv {\rm e}^{\int_x [-S]},
\end{equation}
(where $\, \int_x \,\,\mbox{\rm is the inverse operator of}\,\, D_x\,, \mbox{\rm i.e.}, \, \int_x\,D_x = D_x\,\int_x = \mbox{\boldmath $1$}$) defines a symmetry of the {\em 2ODE (\ref{2oder1})} in the evolutionary form.
\end{cor}
\noindent
{\bf Proof of Corollary \ref{connectSyS}:} Let $X_e$ be the symmetry in the evolutionary form ($X_e = \nu\,\partial_y$). So, its first extension is given by
\begin{equation}
\label{csysproof}
X^{(1)}_e = \nu\,\partial_y+D_x[\nu]\,\partial_z = {\rm e}^{\int_x [-S]}\,\partial_y + {\rm e}^{\int_x [-S]}\,D_x\!\!\left[ \int_x [-S] \right] \!\partial_z =
{\rm e}^{\int_x [-S]}\!\left(\partial_y -S\,\partial_z \right)\!. \nonumber
\end{equation}
Let $I$ be the first integral of the 2ODE (\ref{2oder1}) that is associated with $S$. Then $\,X^{(1)}_e[I] = {\rm e}^{\int_x [-S]} \left(I_y -S\,I_z \right) = {\rm e}^{\int_x [-S]} \left(I_y -\frac{I_y}{I_z}\,I_z \right) = 0. \,\,\,\Box$

\bigskip

The Theorem \ref{connectSSy} and Corollary \ref{connectSyS} are important for establishing a connection between $S$-functions and Lie symmetries. This connection will allow us to develop a method that avoids the use of Darboux polynomials in the process of searching for first integrals for the 2ODE (\ref{2oder1}). The main idea behind the method is based on the concept of associated 1ODE\footnote{This concept was developed in \cite{Nosjmp2009}, page 222.}, which is a 1ODE that has its general solution defined by one of the first integrals of the 2ODE.

\begin{defin}
\label{1odeass}
Let $I$ be a first integral of the {\em 2ODE (\ref{2oder1})} and let $S(x,y,z)$ be the $S$-function associated with {\em (\ref{2oder1})} through $I$. The first order ordinary differential equation defined by
\begin{equation}
\label{1odeassdefs}
\frac{dz}{dy}= - S(x,y,z),
\end{equation}
where $x$ is taken as a parameter, is called {\bf 1ODE$_{\mathbf{[1]}}$ associated} with {\em (\ref{2oder1})} through $I$.
\end{defin}

\begin{teor}
\label{sol1odeass}
Let $I$ be a first integral of the {\em 2ODE (\ref{2oder1})} and let $S(x,y,z)$ be the $S$-function associated with {\em (\ref{2oder1})} through $I$. Then $I(x,y,z)=C$ is a general solution of the {\em 1ODE$_{[1]}$ associated} with {\em (\ref{2oder1})} through $I$.
\end{teor}
\noindent
{\bf Proof of Theorem \ref{sol1odeass}:} The operator defined by $D_a := \partial_y - S\,\partial_z$ annihilates the solutions of the 1ODE$_{[1]}$ associated with the 2ODE (\ref{2oder1}), i.e., the solutions of $\frac{dz}{dy}=-S(x,y,z)$. But $(\partial_y-S\,\partial_z)[I] = I_y - S\,I_z = I_y - \frac{I_y}{I_z}\,I_z \Rightarrow D_a[I] = 0. \,\,\,\Box$

\begin{obs}
\label{solnotinv}
Note that Theorem {\em \ref{sol1odeass}} does not imply that, if we solve the {\em 1ODE (\ref{1odeassdefs})}, we would obtain $I(x,y,z)=C$. The reason is that the variable $x$ (the independent variable of the {\em 2ODE (\ref{2oder1})}) is just a parameter in the {\em 1ODE (\ref{1odeassdefs})}.
\end{obs}
\begin{obs}
\label{solrelati}
Since any function of $x$ is an invariant for the operator  $D_a$, i.e., $D_a[F_1(x)] = (\partial_y - S\,\partial_z)[F_1(x)] = 0$, the relation between a general solution $H(x,y,z)=K$ of the {\em 1ODE (\ref{1odeassdefs})} and the first integral $I(x,y,z)$ of the {\em 2ODE (\ref{2oder1})} is given by $\,I(x,y,z)={\cal F}\left(x,H\right)$, such that the function $\,{\cal F}\,$ satisfies
\begin{equation}
\label{calFeq}
D_x[I] = \frac{\partial {\cal F}}{\partial x} +\left(\frac{\partial H}{\partial x} + z\,\frac{\partial H}{\partial y} + \phi\,\frac{\partial H}{\partial z}\right)  \frac{\partial {\cal F}}{\partial H} = 0.
\end{equation}
\end{obs}

\bigskip

The fact that the $S$-function is related to the symmetry $\nu$ and that the first integral $I$ of the 2ODE (\ref{2oder1}) defines the general solution of the 1ODE$_{[1]}$ becomes clearer if we look at the differential of the first integral $I$ when written in terms of $R$ and $S$: $\,dI = R\,\left[(\phi\,+z\,S)dx - S\,dy - dz\right]$. We can see that, since $\,I_y = -R\,S\,$ and $\,I_z = -R$, if we take $x$ as a parameter then $R$ is also an integrating factor for the 1-form $\,\gamma_a \equiv - S\,dy - dz$, i.e., $dI = R \gamma_a = R\,\left[- S\,dy - dz\right]$ (where $x$ is taken as a parameter). In this way it is clear that $I(x,y,z)=C$ solves the 1ODE given by $dI = R \gamma_a = R\,\left[- S\,dy - dz\right] = 0  \,\,\,\Rightarrow \,\,\, \frac{dz}{dy}=-S$. The reason behind it is that the symmetry of the 2ODE (defined by $\nu$) is in the evolutionary form, i.e., the infinitesimal that multiplies $\partial_x$ is zero (which means that $x$ is an invariant of the symmetry group defined by the generator $\nu \partial_y$). In this way, we could apply the same reasoning to the variable $y$ or to the variable $z$. For example, looking again at $dI$ we see that $\,I_x = R\,(\phi\,+z\,S)\,$ and $\,I_z = -R$, implying that if we now take $y$ as a parameter, then $R$ is also an integrating factor for the 1-form $\,\gamma_b \equiv (\phi\,+z\,S)\,dx - dz$. Now it is clear that $I(x,y,z)=C$ also solves de 1ODE given by
\begin{equation}
\label{twodim2}
dI = R \gamma_b = R\,\left[(\phi\,+z\,S)\,dx - dz\right] = 0  \,\,\,\Rightarrow \,\,\, \frac{dz}{dx}=(\phi\,+z\,S).
\end{equation}

\begin{obs}
\label{solrelati2}
Since any function of $y$ is an invariant for the operator  $D_b$, i.e., $D_b[F_2(y)] = (\partial_x - S_2\,\partial_z)[F_2(y)] = 0$, the relation between a general solution $H_2(x,y,z)=K_2$ of the {\em 1ODE (\ref{twodim2})} and the first integral $I(x,y,z)$ of the {\em 2ODE (\ref{2oder1})} is given by $\,I(x,y,z)={\cal G}\left(y,H_2\right)$, such that the function $\,{\cal G}\,$ satisfies
\begin{equation}
\label{calGeq}
D_x[I] = z\,\frac{\partial {\cal G}}{\partial y} + \left(\frac{\partial H_2}{\partial x} + z\,\frac{\partial H_2}{\partial y} + \phi\,\frac{\partial H_2}{\partial z}\right) \frac{\partial {\cal G}}{\partial H_2} = 0.
\end{equation}
\end{obs}

Since the 1ODE (\ref{twodim2}) can be written as
\begin{equation}
\label{odeass2inv}
\frac{dz}{dx}=-\frac{I_x}{I_z}
\end{equation}
(analogously to $\frac{dz}{dy}=-S$ that could be written as $\frac{dz}{dy}= - \frac{I_y}{I_z}$), we can interpret $S_2 \equiv \frac{I_x}{I_z}$ as `another type' of $S$-function associated with the 2ODE (\ref{2oder1}) through the first integral $I$. And, since $\phi$ is given by $\phi=-\frac{I_x+z\,I_y}{I_z}$ we can substitute $S=-\frac{\phi + S_2}{z}$ in the PDE $D_x[S]=S^2+\phi_z\,S-\phi_y$ and obtain
\begin{equation}
\label{assfun2}
D_x[S_2]=-\frac{1}{z}{S_2}^2+\left(\phi_z-\frac{\phi}{z}\right)\,S_2-\phi_x.
\end{equation}
The 1PDE (\ref{assfun2}) can be considered formally (over the solutions of the 2ODE) as another Riccati 1ODE. So, if we make the transformation
\begin{equation}
\label{stomu}
S_2=z\,\frac{D_x[\mu]}{\mu},
\end{equation}
that is the formal analogous of the transformation (\ref{ricctol2ode}), we get
\begin{equation}
\label{sycon2}
z\,D^2_x[\mu]+(2\,\phi-z\,\phi_z)\,D_x[\mu]+\phi_x\,\mu=0
\end{equation}
which is precisely the symmetry condition for the operator $\,\mu\,\partial_x\,$ to define a symmetry for the 2ODE (\ref{2oder1}).
Finally, making use of the pair $\,I_x = R\,(\phi\,+z\,S)\,$ and $\,I_y = -R\,S\,$ (considering $z$ as a parameter), $I(x,y,z)=C$ also solves de 1ODE given by $\frac{dy}{dx}=-\frac{I_x}{I_y}=\frac{\phi\,+z\,S}{S}$. Defining $S_3 \equiv \frac{I_x}{I_y}$ we can see (after some algebra) that it obeys the 1PDE:
\begin{equation}
\label{assfun3}
D_x[S_3]=-\frac{\phi_y}{\phi}{S_3}^2+\frac{\phi_x-z\,\phi\-y}{\phi}\,S_3+z\,\phi_x,
\end{equation}
that (again over the solutions of the 2ODE) can be viewed as a Riccati 1ODE and so on.

\begin{obs}
\label{solrelati3}
Since any function of $z$ is an invariant for the operator  $D_c \equiv \partial_x - S_3\,\partial_y$, i.e., $D_c[F_3(z)] = (\partial_x - S_3\,\partial_y)[F_3(z)] = 0$, the relation between a general solution $H_3(x,y,z)=K_3$ of the {\em 1ODE} given by $\frac{dy}{dx}=-\frac{I_x}{I_y}=\frac{\phi\,+z\,S}{S}$ and the first integral $I(x,y,z)$ of the {\em 2ODE (\ref{2oder1})} is given by $\,I(x,y,z)={\cal H}\left(z,H_3\right)$, such that the function $\,{\cal H}\,$ satisfies
\begin{equation}
\label{calHeq}
D_x[I] = \phi\,\frac{\partial {\cal H}}{\partial z} + \left(\frac{\partial H_3}{\partial x} + z\,\frac{\partial H_3}{\partial y} + \phi\,\frac{\partial H_3}{\partial z}\right) \frac{\partial {\cal H}}{\partial H_3} = 0.
\end{equation}
\end{obs}

We can, using these results, generalize the concepts of $S$-function and associated 1ODE:

\begin{defin}
\label{sfuns}
Let $I$ be a first integral of the {\em 2ODE (\ref{2oder1})}. The functions defined by $S_{k} := I_{x_i}/I_{x_j}\,$ where $i,j,k \in \{1,2,3\},\,i<j,\,k \notin \{i,j\},\,x_1=x,\,x_2=y,\,x_3=z$, are called \mbox{\boldmath $S$}{\bf -functions} associated with the {\em 2ODE (\ref{2oder1})} through the first integral $I$.
\end{defin}

\begin{defin}
\label{odeasss}
Let $I$ be a first integral of the {\em 2ODE (\ref{2oder1})} and let $S_{k}\,(k=1,2,3)$ be the $S$-functions associated with {\em (\ref{2oder1})} through $I$. The 1ODEs defined by
\begin{equation}
\label{odeassdefs}
\frac{dx_j}{dx_i}= - S_{k},
\end{equation}
where $i,j,k \in \{1,2,3\},\,i<j,\,k \notin \{i,j\},\,x_1=x,\,x_2=y,\,x_3=z$ and $x_k$ is taken as a parameter, are called {\bf 1ODEs associated}  (1ODE$_{\mathbf{[k]}} ,\, (k=1,2,3)\,$) with the {\em 2ODE (\ref{2oder1})} through $I$.
\end{defin}

\begin{defin}
\label{hfuns}
Let $H_i(x,y,z) = K_i$, where $K_i$ is a constant, be a general solution of the associated 1ODE$_{[i]}$. The functions $H_i$ are called \mbox{\boldmath $H$}{\bf -functions}.
\end{defin}

\begin{teor}
\label{sol1odeasss}
Let $I$ be a first integral of the {\em 2ODE (\ref{2oder1})} and let $S_{k}\,(k=1,2,3)$ be the $S$-functions associated with {\em (\ref{2oder1})} through $I$. Then $I(x,y,z)=C$ is a general solution of the {\em 1ODEs} associated with the {\em 2ODE (\ref{2oder1})} through $I$.
\end{teor}
\noindent
{\bf Proof of Theorem \ref{sol1odeasss}:} Suppose that the hypotheses of the theorem are satisfied. Then the operator defined by $D_{[k]} := \partial_{x_i} - S_{k}\,\partial_{x_j}$, where $i,j,k \in \{1,2,3\},\,i<j,\,k \notin \{i,j\},\,x_1=x,\,x_2=y,\,x_3=z$, should annihilate the solutions of the 1ODEs associated with the 2ODE (\ref{2oder1}), i.e., the solutions of $\frac{dx_j}{dx_i}= - S_{k}$. But $( \partial_{x_i} - S_{k}\,\partial_{x_j})[I] = I_{x_i} - S_k\,I_{x_j} = I_{x_i} - \frac{I_{x_i}}{I_{x_j}}\,I_{x_j} \Rightarrow D_{[k]}[I] = 0. \,\,\,\Box$

\begin{obs}
\label{hfunsob}
From Definition \ref{hfuns} and Theorem \ref{sol1odeasss} it follows directly that the first integral $I(x,y,z)$ is a $H$-function.
\end{obs}

\bigskip

In the next section the Theorem \ref{sol1odeasss} and the results described in this section will help us in the construction of a method to find first integrals of rational 2ODEs.

\section{A Method to search for first integrals of a 2ODE}
\label{Iof2odes}
\hspace\parindent
The results of the previous section will allow for the construction of a method that can (in many cases) determine a Liouvillian first integral of a rational 2ODE:
\begin{itemize}
\item First, we will show that, if the integrating factor $R$ and the derivatives of the first integral $I$ have a specific form, we can construct a fast algorithm to determine the $S$-functions associated with the rational 2ODE.
\item In the second subsection, we are going to show the inner works of the method by applying the algorithm to a particular example. Each step will be explained carefully.
\end{itemize}

\subsection{The algorithm}
\label{talgo}
\hspace\parindent

We begin this section with two theorems:

\subsubsection{Two usefull results}
\label{tur}
\hspace\parindent

We will show that

\begin{enumerate}
\item If the integrating factor $R$ for the 1-form $\,\gamma := (\phi\,+z\,S)dx - S\,dy - dz\,$ is a Darboux function, i.e., if $R$ has the format
$\,\exp^{(A/B)}\, \prod_i p_i^{n_i}$, where the $A$, $B$ and the $p_i$ are polynomials in $(x,y,z)$, then the $S$-function associated with the rational 2ODE $z'=\phi(x,y,z)$ through the first integral $I=\int R \gamma$ is rational.
\item If the derivatives of the first integral $I$ are of the format $\,I_x=R\,Q,\, I_y=R\,P,\,I_z=R\,N,\,$ where $N,\,P,\,Q$ are polynomials in $(x,y,z)$, we can reduce the determination of the $S$-functions to the computation of a single polynomial.
\end{enumerate}

\begin{teor}
\label{RDarSrat}
Let $z'=\phi(x,y,z)$ (as in {\em (\ref{2oder1})}) be a rational {\em 2ODE} and let $R=\exp^{(A/B)}\, \prod_i p_i^{n_i}$ be the integrating factor of the 1-form $\,\gamma := (\phi\,+z\,S)dx - S\,dy - dz\,$, where the $A$, $B$ and the $p_i$ are polynomials in $(x,y,z)$, the $p_i$ are irreducible and the $n_i$ are constants. If $S$ is the $S$-function $S_1$ associated with the rational 2ODE $z'=\phi(x,y,z)$ through the first integral $I= \int dI = \int R \gamma$, then $S_1$ is a rational function of $(x,y,z)$.
\end{teor}

\noindent
{\bf Proof of Theorem \ref{RDarSrat}:} From the hypotheses of the theorem we have that $R=\exp^{(A/B)}\, \prod_i p_i^{n_i}$ and therefore

\begin{eqnarray}
\frac{D_x[R]}{R} &=& \frac{ e^{\frac{A}{B}}\,D_x\left[ \prod_i p_i^{n_i} \right] + e^{\frac{A}{B}}\,D_x\left[ \frac{A}{B} \right] \, \prod_i p_i^{n_i} }{e^{\frac{A}{B}}\, \prod_i p_i^{n_i}} = \nonumber \\
&=&   D_x\left[ \frac{A}{B} \right] + \frac{D_x\left[ \prod_i p_i^{n_i} \right] }{ \prod_i p_i^{n_i}} =  D_x\left[ \frac{A}{B} \right] + \sum_i n_i\,\frac{D_x\left[ p_i \right] }{ p_i}. \label{pteo}
\end{eqnarray}
Since $\,D_x := \partial_x + z\,\partial_y + \phi(x,y,z)\,\partial_z\,$ and $\,\phi\,$ is a rational function of $(x,y,z)$, then, from (\ref{pteo}), we have that $\frac{D_x[R]}{R}$ is a rational function of $(x,y,z)$. So, from (\ref{eqproof1r}), $S_1$ is a rational function of $(x,y,z)$.  $\,\,\Box$

\medskip

\begin{teor}
\label{irprd}
Let $z'=\phi(x,y,z)$ (as in {\em (\ref{2oder1})}) be a rational {\em 2ODE} that presents a Liouvillian first integral $I$. Besides, let $I_x = R\,Q, \, I_y = R\,P, \, I_z = R\,N,\,$ where $R$ is an integrating factor of the 1-form $\,\gamma := N \, [(\phi\,+z\,S)dx - S\,dy - dz]\,$, $\,N,\,P$ and $Q$ are polynomials, $S$ is the $S$-function $S_1$ and $N$ is the denominator of $\phi$. Then $R$ has the format $\exp^{(A/B)}\, \prod_i p_i^{n_i}$, where the $A$, $B$ and the $p_i$ are polynomials in $(x,y,z)$, the $n_i$ are constants and the $S$-functions associated with the rational 2ODE $z'=\phi(x,y,z)$ through the first integral $I$ are given by $S_1 = \frac{P}{N},\, S_2 = \frac{Q}{N},\,$ and $\,S_3 = \frac{Q}{P}$.
\end{teor}

\noindent
{\bf Proof of Theorem \ref{irprd}:} From the hypotheses of the theorem it follows straightforward that the $S$-functions are given by $S_1 = \frac{P}{N},\, S_2 = \frac{Q}{N},\,$ and $\,S_3 = \frac{Q}{P}$. So (by Theorem \ref{sol1odeasss}) $I(x,y,z)=C$ is the general solution of the rational 1ODEs  $\frac{dz}{dy} = -\frac{P}{N},\, \frac{dz}{dx} = -\frac{Q}{N},\,$ and $\,\frac{dy}{dx} = -\frac{Q}{P}$. Then, by the results presented in \cite{Nosjpa2002-1,Nosjpa2002-2,Noscpc2007}, the 1-forms $\gamma_1 \equiv \,P\,dy+N\,dz,\,\,\gamma_2 \equiv \,Q\,dx+N\,dz\,$ and $\,\gamma_3 \equiv \,Q\,dx+P\,dy$ present integrating factors of the form: $R_1 = \exp^{(A_1/B_1)}\, \prod_i {p_1}_i^{{n_1}_i}$, $R_2 = \exp^{(A_2/B_2)}\, \prod_i {p_2}_i^{{n_2}_i}$ and $R_3 = \exp^{(A_3/B_3)}\, \prod_i {p_3}_i^{{n_3}_i}$, where ${p_1}_i$, ${A_1}$, and ${B_1}$ are polynomials in $(y,z)$, ${p_2}_i$, ${A_2}$, and ${B_2}$ are polynomials in $(x,z)$ and ${p_3}_i$, ${A_3}$, and ${B_3}$ are polynomials in $(x,y)$. Since $\,dI=I_x dx+I_y dy+I_z dz = R\,(Q\,dx+P\,dy+N\,dz)$, $R$ is also an integrating factor for the associated 1ODEs. This means that $\,R_1={\cal F}_1(I)\,R\,$  (where $\,{\cal F}_1\,$ is a non constant function of $\,I\,$) or $\,R_1=k_1\,R\,$ (where $\,k_1\,$ is a constant). Following the same reasoning we have that $\,R_2={\cal F}_2(I)\,R\,$  (where $\,{\cal F}_2\,$ is a non constant function of $\,I\,$) or $\,R_2=k_2\,R\,$ (where $\,k_2\,$ is a constant), implying that $\,R_1\,$ and $\,R_2\,$ are also integrating factors for the 1-form $\gamma$. So,  $\,R_1={\cal G}(I)\,R_2\,$  (where $\,{\cal G}\,$ is a non constant function of $\,I\,$) or $\,R_1=k\,R_2\,$ (where $\,k\,$ is a constant). Now, we have two cases to consider:
\begin{itemize}
\item First possibility: The 2ODE (\ref{2oder1}) presents an elementary first integral. In this case (see \cite{Nosjmp2009}) there exists an algebraic integrating factor of the form $\,\prod_i p_i^{n_i}\,$ and the theorem is demonstrated.
\item Second possibility: The 2ODE (\ref{2oder1}) does not have an elementary first integral. In this case the only possibility is $\,R_1=k\,R_2\,$ (for, if $\,R_1={\cal G}(I)\,R_2\,$, this would imply that $\,{\cal G}(I)=R_1/R_2\,\Rightarrow\,$ exists an elementary first integral which contradicts the hypothesis) implying that the 1-form $\,\gamma\,$ has an integrating factor of the form $\exp^{(A/B)}\, \prod_i p_i^{n_i}$, where the $A$, $B$ and the $p_i$ are polynomials in $(x,y,z)$.  $\,\,\Box$
\end{itemize}

\bigskip

\begin{obs}
\label{almostgeneral}
Although at first glance it may seem that these conditions on the form of the integrating factor and on the derivatives of the first integral are too restrictive, we will see (see section \ref{performance} below) that, in practice, if we are looking for Liovillian first integrals of rational 2ODEs, we can succeed in a considerable number of cases, making the method of practical relevance.
\end{obs}
\begin{obs}
\label{higheficience}
The computation of the $S$-functions is more efficient than the computation of the Darboux polynomials (or Lie Symmetries) in a great number of cases where the Lie and Darboux methods are difficult to be applied in practice (see section \ref{performance} below).
\end{obs}

\medskip

\subsubsection{Construction of a method}
\label{coam}
\hspace\parindent

In this section, based on the results just presented, we will construc algorithms (semi) to deal with rational 2ODEs restricted to the conditions described in the previous section. Therefore, to begin with, let us assume that the rational 2ODE (\ref{2oder1}) presents a Liouvillian first integral $I$ such that $I_x = R\,Q, \, I_y = R\,P, \, I_z = R\,N,\,$ and $R$ is an integrating factor of the form $\exp^{(A/B)}\, \prod_i p_i^{n_i}$, where $\,N,\,P$, $Q$ $A$, $B$ and the $p_i$ are polynomials in $(x,y,z)$ and $N$ is the denominator of $\phi$. Substituting $\,S_1= \frac{P}{N}\,$ into the 1PDE $\,D_x[S_1]=S_1^2+\phi_{z}\,S_1-\phi_y\,$ we obtain:
\begin{equation}
\label{eqP}
-{P}^{2} - \left( N_x+z\,N_y+M_z \right) P+ D[P] -M\,N_y+M_y\,N=0,
\end{equation}
where $\,D \equiv N\,D_x = N\,\partial_x +z\,N\,\partial_y+M\,\partial_z$. The idea is to construct a polynomial ${\cal P}$ with undetermined coefficients and substitute it in (\ref{eqP}). The result will be an equation of the type $Polynomial = 0$. Then, we collect the polynomial equation in the variables $x,\,y,\,z$ and equate the coefficients of each monomial to zero, obtaining a system of equations. If this system presents a solution we will have found $S_1$.

Analogously, we could substitute  $\,S_2= \frac{Q}{N}\,$ into the 1PDE $\,D_x[S_2]=-\frac{1}{z}{S_2}^2+\left(\phi_z-\frac{\phi}{z}\right)\,S_2-\phi_x\,$ and obtain:
\begin{equation}
\label{eqQ}
Q^2-\left(N_y\,{z}^{2}+ \left(M_z+N_x \right) z -M \right) Q - \left(M\,N_x-M_x\,N -D[Q]\right) z=0.
\end{equation}
As before, we construct a polynomial ${\cal Q}$ with undetermined coefficients and substitute it in (\ref{eqQ}). The result will be another polynomial equation of the type $Polynomial = 0$. Again, we collect the polynomial equation in the variables $x,\,y,\,z$ and  and equate the coefficients of each monomial to zero, obtaining a system of equations. If this system presents a solution we will have found $S_2$.

If we succeed in the determination of $\,S_1\,$ (or $\,S_2\,$), we can construct the associated 1ODE and try to solve it. From the solution of the associated 1ODE we can find the first integral $I(x,y,z)$ by solving the 1ODE obtained from the characteristic system of de 1PDE (\ref{calFeq}) (or of the 1PDE (\ref{calGeq})). So, we have the following algorithms (respectively to find $S_1$ and $S_2$:

\bigskip

\begin{algor}  (AS1)
\label{as1}
	\begin{enumerate}
		\item Let $n_{max}$ =  $\max(degree (M)-1,degree (N))$.
		\item Let $n=0$.
		\item Let $n=n+1$.
		\item if $n>n_{max}$ then FAIL.
             \item Construct the $D_x$ operator.
		\item Construct a generic polynomial ${\cal P}$ of degree $n$ in $(x,y,z)$ with undetermined coefficients $a_i$.
		\item Substitute ${\cal P}$ in equation (\ref{eqP}), collect the resulting polynomial equation in the variables $x,\,y,\,z$ and equate the coefficients of each monomial to zero, obtaining a system ${\cal A}$ of algebraic equations.
		\item Solve the system ${\cal A}$ with respect to $\{a_i\}$. If no solution is found, then go to step 3.
		\item Substitute the solution in ${\cal P}$ (obtaining $P$) and construct $S_1 = P/N$.
		\item Construct the associated 1ODE$_{[1]}$ ($\frac{dz}{dy}=-\frac{P}{N}$) and try to solve it to obtain a solution $H_1(x,y,z)=C_1$. If no solution is found, then go to step 3.
		\item Solve $h=H_1(x,y,z)$ for one of the variables $x$, $y$ or $z$ (or any operand belonging to $H_1$ and to $D_x[H_1]$) and substitute it in the 1PDE (\ref{calFeq}). Then, try to solve the characteristic equation -- a 1ODE in $(x,h)$ -- to obtain ${\cal F}(x,h)=K_1$. If no solution is found, then go to step 3.
		\item Construct the first integral $I={\cal F}(x,H_1)$.
	\end{enumerate}
\end{algor}

\bigskip

\begin{algor}  (AS2)
\label{as2}
	\begin{enumerate}
		\item Let $n_{max}$ =  $\max(degree (M)-1,degree (N))$.
		\item Let $n=0$.
		\item Let $n=n+1$.
		\item if $n>n_{max}$ then FAIL.
             \item Construct the $D_x$ operator.
		\item Construct a generic polynomial ${\cal Q}$ of degree deg in $(x,y,z)$ with undetermined coefficients $b_i$.
		\item Substitute ${\cal Q}$ in equation (\ref{eqQ}), collect the resulting polynomial equation in the variables $x,\,y,\,z$ and equate the coefficients of each monomial to zero, obtaining a system ${\cal B}$ of algebraic equations.
		\item Solve the system ${\cal B}$ with respect to $\{b_i\}$. If no solution is found, then go to step 3.
		\item Substitute the solution in ${\cal Q}$ (obtaining $Q$) and construct $S_2 = Q/N$.
		\item Construct the associated 1ODE$_{[2]}$ ($\frac{dz}{dx}=-\frac{Q}{N}$) and try to solve it to obtain a solution $H_2(x,y,z)=C_2$. If no solution is found, then go to step 3.
		\item Solve $h=H_2(x,y,z)$ for one of the variables $x$, $y$ or $z$ (or any operand belonging to $H_2$ and to $D_x[H_2]$) and substitute it in the 1PDE (\ref{calGeq}). Then, try to solve the characteristic equation -- a 1ODE in $(y,h)$ -- to obtain ${\cal G}(y,h)=K_2$. If no solution is found, then go to step 3.
		\item Construct the first integral $I={\cal G}(y,H_2)$.
	\end{enumerate}
\end{algor}

\bigskip

\subsection{The inner works of the method}
\label{iwm}
\hspace\parindent

Here we will show an example of the method in action. For this, we will apply it to a `problematic' 2ODE, i.e., a 2ODE that is very hard to be reduced by a canonical method.

Consider the rational 2ODE given by
\begin{equation}
\label{exmethod}
z'={\frac {{x}^{5}z-{x}^{4}{z}^{2}-3\,z{x}^{4}+4\,{x}^{3}{z}^{2}-xy+xz+yz
-{z}^{2}-y}{{x}^{5}-y}}.
\end{equation}

We see that $M={x}^{5}z-{x}^{4}{z}^{2}-3\,z{x}^{4}+4\,{x}^{3}{z}^{2}-xy+xz+yz-{z}^{2}-y$ and $N={x}^{5}-y$. Therefore, the maximum degree of the polynomial $P$ is $5$ and $D_x = \partial_x+z \partial_y+(M/N)\partial_z$. Let's begin with $n=1$ leading to ${\cal P}=a_0+a_1\,x+a_2\,y+a_3\,z$. Substituting $P={\cal P}$ in equation (\ref{eqP}) and collecting with respect to $(x,y,z)$ we get
\begin{eqnarray}
&&\left( {\it a_1}+1 \right) {x}^{6}+{\it a_2}\,y{x}^{5}+ \left( -2\,
{\it a_1}-{\it a_2}-2 \right) z{x}^{5}+ \left( {\it a_0}+{\it a_1}
+1 \right) {x}^{5} \nonumber \\
&&-2\,{\it a_2}\,z{x}^{4}y+2\,{\it a_2}\,{x}^{4}y+
 \left( -{\it a_3}+1 \right) {x}^{4}{z}^{2}+ \left( 3+5\,{\it a_3}+8
\,{\it a_1}-2\,{\it a_0} \right) z{x}^{4} \nonumber \\
&&+2\,{\it a_0}\,{x}^{4}+8\,
{\it a_2}\,z{x}^{3}y+ \left( 4\,{\it a_3}-4 \right) {x}^{3}{z}^{2}+8
\,{\it a_0}\,{x}^{3}z+ \left( {{\it a_1}}^{2}+{\it a_1} \right) {x}^{2} \nonumber \\
&&+ \left( 2\,{\it a_2}\,{\it a_1}+{\it a_1}+{\it a_2}+{\it a_3
} \right) xy+ \left( 2\,{\it a_1}\,{\it a_3}-3\,{\it a_1}-1
 \right) xz  \nonumber \\
&&+ \left( 2\,{\it a_0}\,{\it a_1}+{\it a_0} \right) x+
 \left( {{\it a_2}}^{2}+{\it a_2} \right) {y}^{2}+ \left( 2\,{\it
a_3}\,{\it a_2}-2\,{\it a_2} \right) yz+ \left( 2\,{\it a_0}\,{\it a_2} \right. \nonumber \\
&& \left. +{\it a_0}+{\it a_1}+{\it a_3} \right) y+ \left( {{\it
a_3}}^{2}-2\,{\it a_3}+1 \right) {z}^{2}+ \left( 2\,{\it a_0}\,{
\it a_3}-3\,{\it a_0} \right) z+{{\it a_0}}^{2}=0. \nonumber \label{pequa}
\end{eqnarray}
Equating the coefficients of the polynomial equation (\ref{pequa}) to zero, we obtain a system ${\cal A}$ of algebraic equations. The system can be easily solved:
\begin{equation}
\label{solsysA}
{a_0 = 0, a_1 = -1, a_2 = 0, a_3 = 1}.
\end{equation}
This leads to $P=z-x$ and
\begin{equation}
\label{S1ex1}
S_1=\frac{z-x}{x^5-y}.
\end{equation}
The associated 1ODE$_{[1]}$ is
\begin{equation}
\label{1ode1ex1}
\frac{dz}{dy} = - \frac{z-x}{x^5-y},
\end{equation}
whose solution is $z = \left( {x}^{5}-y \right) \, K_1+x$ and, therefore,
\begin{equation}
\label{H1ex1}
K_1 = {\frac {z-x}{{x}^{5}-y}} \,\,\,\Rightarrow\,\,\, H_1={\frac {z-x}{{x}^{5}-y}}.
\end{equation}
At this point we recall that, although $I(x,y,z) = C$ is a solution of the associated 1ODE, the function $H_1$ (that defines the solution $H_1(x,y,z)=K_1$) is not necessarily the first integral $I$. The 1PDE that relates $H_1$ and $I$ is
\begin{equation}
\label{1pdecalFex1}
\frac{\partial {\cal F}}{\partial x} +D_x[H_1] \,  \frac{\partial {\cal F}}{\partial h} = 0 \,\,\,\,\, \Rightarrow \,\,\,\,\, \frac{\partial {\cal F}}{\partial x} +{\frac { \left( -z+x \right)  \left( z{x}^{4}+4\,{x}^{4}-4\,{x}^{3}z-y
 \right) }{ \left( {x}^{5}-y \right) ^{2}}}
 \,  \frac{\partial {\cal F}}{\partial h} = 0. \nonumber
\end{equation}
Solving $h={\frac {z-x}{{x}^{5}-y}}$ for $z$ we obtain $z=h{x}^{5}-hy+x$. Substituting it in the 1PDE we have:
\begin{equation}
\label{1pdecalFex1hx}
\frac{\partial {\cal F}}{\partial x} -h \left( h{x}^{4}-4\,h{x}^{3}+1 \right) \, \frac{\partial {\cal F}}{\partial h} = 0.
\end{equation}
The characteristic equation for the 1PDE (\ref{1pdecalFex1hx}) is
\begin{equation}
\label{1odehx}
\frac{dh}{dx}=-h \left( h{x}^{4}-4\,h{x}^{3}+1 \right),
\end{equation}
whose solution is
\begin{equation}
\label{solhx}
h=\frac{1}{k_1\,{\rm e}^{x}-{x}^{4}} \,\,\,\, \Rightarrow \,\,\,\, {\cal F}(x,h) = {\frac {h \,{x}^{4}+1}{h \, {{\rm e}^{x}}}}.
\end{equation}
We have
\begin{equation}
\label{Iex1}
I = {\cal F}(x,H_1) = {\frac { \left( z{x}^{4}-y \right) {{\rm e}^{-x}}}{z-x}}.
\end{equation}
Finally, solving for $z$, we have $z=\frac{dy}{dx}={\frac {Cx-{{\rm e}^{-x}}y}{-{x}^{4}{{\rm e}^{-x}}+C}}$ whose solution is the general solution of the 2ODE:
\begin{equation}
\label{sol2odeex1}
y = \left( \int \!Cx{{\rm e}^{\int \!{\frac {{{\rm e}
^{-x}}}{-{x}^{4}{{\rm e}^{-x}}+C}}{dx}}} \left( -{x}^{4}{{\rm e}^{-x}}
+C \right) ^{-1}{dx}+K \right) {{\rm e}^{\int \!-{\frac {{
{\rm e}^{-x}}}{-{x}^{4}{{\rm e}^{-x}}+C}}{dx}}}.
\end{equation}

\bigskip

\section {The {\it InSyDE} Package}
\label{insyde}

In this section we will present a Maple implementation of the algorithms {\it AS1} and {\it AS2}.

\bigskip

{\bf Summary of the commands:}

\begin{itemize}
\item {\tt Dx} constructs the $D_x$ operator associated with the 2ODE.

\item {\tt Sfunction} tries to determine a $S$-function associated with the 2ODE.

\item {\tt Exodes} determines the 1ODEs associated with the 2ODE (from the knowledge of a $S$-function).

\item {\tt Hfunction} tries to find the general solution for an associated 1ODE.

\item {\tt PDEassol} constructs and tries to solve the 1PDEs that relate the first integral $I$ with the $H$-functions.

\item {\tt Invade} tries to determine a Liouvillian first integral of the 2ODE from solving the associated 1PDE.


\item {\tt Gensol} tries to find the general solution of the 2ODE (from the knowledge of the first integral $I$).

\end{itemize}

\subsection {Package commands}
\label{packcom}

Here we present a detailed description of the package's commands\footnote{This subsection and the next one may contain some information already presented in the previous sections; this is necessary to produce a self-contained description of the package.}.

\noindent
\subsubsection{Command name: {\tt Dx}}
\label{dxcom}

\noindent {\it Feature:} This command constructs the $D_x$ operator.
\bigskip

\noindent
{\it Calling sequence:}\footnote{In what follows the {\it input} can be recognized by the Maple prompt {\tt [>}.}

\begin{verbatim}
[> Dx(ode);
\end{verbatim}

\noindent
{\it Parameters:}
\medskip

\hspace
\parindent
{\tt ode} - The rational 2ODE.

\bigskip

\noindent
{\it Synopsis:}
\smallskip
\smallskip

The command {\tt Dx} returns the operator $D_x \equiv \partial_x+z\partial_y+\phi\partial_z$. This operator calculates the total derivative $\frac{d}{dx}$ (of any function of $(x,y,z)$) over the solutions of the 2ODE $\,z'=\phi(x,y,z)$.

\bigskip

%%%%%%%
\noindent
\subsubsection{Comando: {\tt Sfunction}}
\label{sfuncom}

\noindent {\it Feature:} This command tries to find an $S$-function associated to the rational 2ODE.
\bigskip

\noindent
{\it Calling sequence:}

\begin{verbatim}
[> Sfunction(ode);
\end{verbatim}

\noindent
{\it Parameters:}
\medskip

\hspace
\parindent
{\tt ode} - The rational 2ODE.
\medskip

\bigskip

\noindent
{\it Extra parameters:}
\medskip

\hspace
\parindent
{\tt Sn = ns} - Where {\tt ns} $\in \, \{1,2,3,4\}$ denotes if we are looking for $S_1$, $S_2$ or $S_3$ or all of them (value = 4). The default is 1.
\medskip

\hspace
\parindent
{\tt Deg = n} - Where {\tt n} is a positive integer denoting the degree of the polynomial $P$ (or $Q$). The default is 1.
\medskip

\hspace
\parindent
{\tt Den = deno} - Where {\tt deno} is the denominator of the $S$-function.
\bigskip


\noindent
{\it Synopsis:}
\smallskip
\smallskip

The command {\tt Sfunction} tries to find a $S$-function associated with the rational 2ODE through a Liouvillian first integral $I$. The command computs (if possible) a polynomial $P$ (or $Q$) that is the numerator of $S_1$ (or $S_2$). The $S$-functions are the basis for the algorithms implemented here. From one of them the command can compute the others. As we have a maximum degree for the polynomials $P$ and $Q$, we can use the parameter {\tt Deg} to inform the program of the degree we will use. We have also the parameter {\tt Den} that can acelerate the process by indicating the denominator of the $S$-function.

\bigskip


%%%%%%%
\noindent
\subsubsection{Comando: {\tt Exodes}}
\label{exodcom}

\noindent {\it Feature:} This command determines the associated 1ODEs.
\bigskip

\noindent
{\it Calling sequence:}

\begin{verbatim}
[> Exodes(ode);
\end{verbatim}

\noindent
{\it Parameters:}
\medskip

\hspace
\parindent
{\tt ode} - The rational 2ODE.

\bigskip

\noindent
{\it Extra parameters:}
\medskip

\hspace
\parindent
{\tt Sn = ns} - Where {\tt ns} $\in \, \{1,2,3\}$ denotes if we are going to use $S_1,\,S_2$ or $S_3$. The default is 1.
\medskip

\hspace
\parindent
{\tt En = ne} - Where {\tt ne} $\in \, \{1,2,3\}$ denotes if we are looking for 1ODE$_{[1]}$, 1ODE$_{[2]}$ or 1ODE$_{[3]}$. The default is 1.
\medskip

\hspace
\parindent
{\tt Deg = n} - Where {\tt n} is a positive integer denoting the degree of the polynomial $P$ (or $Q$). The default is 1.
\medskip

\hspace
\parindent
{\tt Den = deno} - Where {\tt deno} is the denominator of the $S$-function.
\medskip

\hspace
\parindent
{\tt Sfun = S1} - Where {\tt S1} is a $S$-function $S_1$.
\bigskip

\noindent
{\it Synopsis:}
\smallskip
\smallskip

The command {\tt Exodes} uses the $S$-functions to construct the associated 1ODEs. Since $\,\phi=\frac{M}{N}=\frac{I_x+zI_y}{I_z}=S_2+z\,S_1\,$ and $\,S_3=\frac{S_2}{S_1}\,$, from a single $S$-function we can determine the others. Therefore from a single $S$-function we can find the three associated 1ODEs. The first three extra parameters are the same of the {\tt Sfunction} command because we have to calculate it before find the associated 1ODEs. The last extra parameter ({\tt Sfun = S1}) allows the user to pass the $S$-function $S_1$ to the {\tt Exodes} command.

\bigskip


%%%%%%%
\noindent
\subsubsection{Comando: {\tt Hfunction}}
\label{hfuncom}

\noindent {\it Feature:} This command tries to find the solutions of the associated 1ODEs.
\bigskip

\noindent
{\it Calling sequence:}

\begin{verbatim}
[> Hfunction(ode);
\end{verbatim}

\noindent
{\it Parameters:}
\medskip

\hspace
\parindent
{\tt ode} - The rational 2ODE.

\bigskip

\noindent
{\it Extra parameters:}
\medskip

\hspace
\parindent
{\tt Sn = ns} - Where {\tt ns} $\in \, \{1,2,3\}$ denotes if we are going to use $S_1,\,S_2$ or $S_3$. The default is 1.
\medskip

\hspace
\parindent
{\tt En = ne} - Where {\tt ne} $\in \, \{1,2,3,4\}$ denotes if we are looking for $H_1$, $H_2$ or $H_3$. The default is 1.
\medskip

\hspace
\parindent
{\tt Deg = n} - Where {\tt n} is a positive integer denoting the degree of the polynomial $P$ (or $Q$). The default is 1.
\medskip

\hspace
\parindent
{\tt Den = deno} - Where {\tt deno} is the denominator of the $S$-function.
\bigskip

\hspace
\parindent
{\tt Sfun = S1} - Where {\tt S1} is a $S$-function $S_1$.
\bigskip

\noindent
{\it Synopsis:}
\smallskip
\smallskip

The command {\tt Hfunction} tries to solve the associated 1ODEs. As the $H$-functions are not necessarily the first integral $I$, there is still a need to solve a 1ODE which is the characteristic equation of the 1PDE that relates the (specified) $H$-function with $I$.
\bigskip

\noindent
\subsubsection{Comando: {\tt PDEassol}}
\label{pdeascom}

\noindent {\it Feature:} This command constructs and tries to solve the 1PDEs that relates the $H$-functions to the first integral $I$.
\bigskip

\noindent
{\it Calling sequence:}

\begin{verbatim}
[> PDEassol(ode);
\end{verbatim}

\noindent
{\it Parameters:}
\medskip

\hspace
\parindent
{\tt ode} - The rational 2ODE.

\bigskip

\noindent
{\it Extra parameters:}
\medskip

\hspace
\parindent
{\tt Sn = ns} - Where {\tt ns} $\in \, \{1,2,3\}$ denotes if we are going to use $S_1,\,S_2$ or $S_3$. The default is 1.
\medskip

\hspace
\parindent
{\tt En = ne} - Where {\tt ne} $\in \, \{1,2,3,4\}$ denotes if we are looking for $H_1$, $H_2$ or $H_3$. The default is 1.
\medskip

\hspace
\parindent
{\tt Deg = n} - Where {\tt n} is a positive integer denoting the degree of the polynomial $P$ (or $Q$). The default is 1.
\medskip

\hspace
\parindent
{\tt Den = deno} - Where {\tt deno} is the denominator of the $S$-function.
\medskip

\hspace
\parindent
{\tt Sfun = S1} - Where {\tt S1} is a $S$-function $S_1$.
\bigskip


\noindent
{\it Synopsis:}
\smallskip
\smallskip

The command {\tt PDEassol} constructs the associated 1PDEs and tries to solve them. The command generates also the 1ODE which is the characteristic equation of the associated 1PDE and its solution (if possible).
\bigskip

%%%%%%%
\noindent
\subsubsection{Comando: {\tt Invade}}
\label{invcom}

\noindent {\it Feature:} This command tries to find a Liouvillian first integral of the rational 2ODE.
\bigskip

\noindent
{\it Calling sequence:}

\begin{verbatim}
[> Invade(ode);
\end{verbatim}

\noindent
{\it Parameters:}
\medskip

\hspace
\parindent
{\tt ode} - The rational 2ODE.

\bigskip

\noindent
{\it Extra parameters:}
\medskip

\hspace
\parindent
{\tt Sn = ns} - Where {\tt ns} $\in \, \{1,2,3\}$ denotes if we are going to use $S_1,\,S_2$ or $S_3$. The default is 1.
\medskip

\hspace
\parindent
{\tt En = ne} - Where {\tt ne} $\in \, \{1,2,3\}$ denotes if we are looking for $H_1$, $H_2$ or $H_3$. The default is 1.
\medskip

\hspace
\parindent
{\tt Deg = n} - Where {\tt n} is a positive integer denoting the degree of the polynomial $P$ (or $Q$). The default is 1.
\medskip

\hspace
\parindent
{\tt Den = deno} - Where {\tt deno} is the denominator of the $S$-function.
\medskip

\hspace
\parindent
{\tt Sfun = S1} - Where {\tt S1} is a $S$-function $S_1$.
\bigskip


\noindent
{\it Synopsis:}
\smallskip
\smallskip

The command {\tt Invade} tries to find a Liouvillian first integral of the rational 2ODE through the determination of a $S$-function. This routine coordenates the whole process. Once we have the solution of the associated 1PDE the construction of the first integral $I$ is strightforward.
\bigskip


%%%%%%%
\noindent
\subsubsection{Comando: {\tt Gensol}}
\label{pdeascom}

\noindent {\it Feature:} This command determines (if possible) the general solution of the rational 2ODE.
\bigskip

\noindent
{\it Calling sequence:}

\begin{verbatim}
[> Gensol(ode);
\end{verbatim}

\noindent
{\it Parameters:}
\medskip

\hspace
\parindent
{\tt ode} - The rational 2ODE.

\bigskip

\noindent
{\it Extra parameters:}
\medskip

\hspace
\parindent
{\tt Sn = ns} - Where {\tt ns} $\in \, \{1,2,3\}$ denotes if we are going to use $S_1,\,S_2$ or $S_3$. The default is 1.
\medskip

\hspace
\parindent
{\tt En = ne} - Where {\tt ne} $\in \, \{1,2,3\}$ denotes if we are looking for $H_1$, $H_2$ or $H_3$. The default is 1.
\medskip

\hspace
\parindent
{\tt Deg = n} - Where {\tt n} is a positive integer denoting the degree of the polynomial $P$ (or $Q$). The default is 1.
\medskip

\hspace
\parindent
{\tt Den = deno} - Where {\tt deno} is the denominator of the $S$-function.
\medskip

\hspace
\parindent
{\tt Sfun = S1} - Where {\tt S1} is a $S$-function $S_1$.
\bigskip


\noindent
{\it Synopsis:}
\smallskip
\smallskip

The command {\tt Gensol} determines (if possible) a general solution for the rational 2ODE. If the {\tt Invade} command succeeds in finding a first integral $I$ for the 2ODE, the {\tt Gensol} command simply solves $I(x,y,z)=C$ for $z$ and tries to solve the 1ODE $z=\varphi(x,y,C)$.

\bigskip


\begin{obs}
\label{extrapar}
The extra parameters are not mandatory -- some are even redundant. For example, if we are to provide the $S$-function we do not need to provide the degree of the polynomials $P$ (or $Q$). If these redundant parameters are supplied simultaneously, the package will take care of doing (hopefully) the best choice.
\end{obs}


%%%%%%%%%%%%%%%%
\subsection {Example of the usage of the package commands}
\label{exusagecom}

In this section we use two 2ODEs and show the commands in action so that the reader (possible user) can solve the most common doubts by direct observation.

Consider the 2ODE (example presented in the section \ref{iwm})

\begin{equation}
\label{expackode1}
z'={\frac {{x}^{5}z-{x}^{4}{z}^{2}-3\,z{x}^{4}+4\,{x}^{3}{z}^{2}-xy+xz+yz
-{z}^{2}-y}{{x}^{5}-y}}.
\end{equation}
%
and let's supose that we want to solve / study it. After opening a Maple session, we will load the required packages:
%
\begin{verbatim}
[> with(DEtools):  read(`InSyDE.txt`):
\end{verbatim}
%
The {\tt :} sign after the command line avoids printing (on-screen) of the result. The {\it DEtools} package loads several commands to handle ODEs. The {\tt read (`InSyDE.txt`):} command loads our package. Let's `enter' the 2ODE (\ref{expackode1}) by typing
%
\begin{verbatim}
[> _2ode := diff(y(x),x,x) = (x^5*(diff(y(x),x))-x^4*(diff(y(x)
   ,x))^2-3*(diff(y(x),x))*x^4+4*x^3*(diff(y(x),x))^2-x*y(x)+x*
   (diff(y(x),x))+y(x)*(diff(y(x),x))-(diff(y(x),x))^2-y(x))/(x
   ^5-y(x)):
\end{verbatim}
%
Let us start by searching a $S$-function. Typing
%
\begin{verbatim}
[> S1 := Sfunction(_2ode);
\end{verbatim}
%
results in the output
%
\begin{equation}
S1 := -{\frac {-z+x}{{x}^{5}-y}}
\end{equation}		
%
From $S_1$ we can determine the associated 1ODEs. By typing
\begin{verbatim}
[> _1odeas := Exodes(_2ode,En=4);
\end{verbatim}
we get
\begin{eqnarray}
&&\_1odeas :={\frac {d}{dy}}z \left( y \right) ={\frac {-z \left( y \right) +x}{{x
}^{5}-y}},  \nonumber \\
&& {\frac {d}{dx}}z \left( x \right) ={\frac {{x}^{5}z \left( x
 \right) -{x}^{4}  z \left( x \right)^{2}-3\,z \left(
x \right) {x}^{4}+4\,{x}^{3} z \left( x \right)^{2}-x
y+yz \left( x \right) -y}{{x}^{5}-y}}, \nonumber \\
&&  {\frac {d}{dx}}y \left( x \right) =-{\frac {{x}^{5}z-{x}^{4}{z}^{2}-3\,z{x}^{4}+4\,{x}^{3}{z}^{
2}-xy \left( x \right) +y \left( x \right) z-y \left( x \right) }{-z+x}} \nonumber
\end{eqnarray}		
%
We can try solving the associated 1ODE$_{[1]}$ with
\begin{verbatim}
[> H1 := Hfunction(_2ode);
\end{verbatim}
that leads to the output
\begin{equation}
H1 := -\frac {-z+x}{{x}^{5}-y}
\end{equation}		
Once we have found the $H$-function $H_1$, we can apply the {\tt PDEassol} command in order to determine the functional relation between $I$ and $H_1$. The application of the command
\begin{verbatim}
[> pdeassol := PDEassol(_2ode);
\end{verbatim}
%
results in
\begin{eqnarray}
pdeassol &:=& \left[\frac{\partial F(x,h)}{\partial x} -h \left( h{x}^{4}-4\,h{x}^{3}+1 \right)\,\frac{\partial F(x,h)}{\partial h}=0, \right. \nonumber \\
&& {\frac {d}{dx}}h \left( x \right) =-h \left( x \right)  \left( h \left( x \right) {x}^{4}-4\,h \left( x \right) {x}^{3}+1 \right), \nonumber \\
&& \left. {\it \_H}_{{1}}=-{\frac {-z+x}{{x}^{5}-y}},F \left( x,h \right) ={\frac {h{x}^{4}+1}{{{\rm e}^{x}}h}} \right] \nonumber
\end{eqnarray}		
%
From the solution of the associated PDE$_{[1]}$, we can find a first integral $I$. Using the command
\begin{verbatim}
[>  Inv := Invade(_2ode);
\end{verbatim}
we find
\begin{equation}
Inv := -{\frac { \left( z{x}^{4}-y \right) {{\rm e}^{-x}}}{-z+x}}
\end{equation}
We can try to completely solve the 2ODE by using the command
\begin{verbatim}
[> sol2ode := Gensol(_2ode);
\end{verbatim}
%
This leads to the following output
\begin{equation}
sol2ode := y \left( x \right) = \left( \int \!- \frac{x{\it \_C1}\,{{\rm e}^{-\int \!{
\frac {{{\rm e}^{-x}}}{{x}^{4}{{\rm e}^{-x}}-{\it \_C1}}}{dx}}}}
{{x}^{4}{{\rm e}^{-x}}-{\it \_C1}}{dx}+{\it \_C2}
 \right) {{\rm e}^{\int \!{\frac {{{\rm e}^{-x}}}{{x}^{4}{{\rm e}^{-x}
}-{\it \_C1}}}{dx}}}
\end{equation}
%
%Finally, with can calculate an integrating factor and a symmetry of the 2ODE:
%\begin{verbatim}
%[> infac := Infact(_2ode);
%[> symm := Symmeo(_2ode);
%\end{verbatim}
%
%\begin{eqnarray}
%{\rm {\it infac}} &:=& {\frac {{{\rm e}^{-x}}}{ \left( -z+x \right) ^{2}}} \nonumber \\
%{\rm {\it symm}} &:=& \left[ {{\rm e}^{\int \!L \left( x,y,z \right) {dx}}},L \left( x,y,z
% \right) ={\frac {-z+x}{{x}^{5}-y}} \right] \nonumber
%\end{eqnarray}
%

\bigskip

Now, consider the following 2ODE:
\begin{equation}
\label{expackode2}
z'=-{\frac {{x}^{2}{z}^{8}+y{z}^{4}x-zx+y}{x{z}^{2} \left( 3\,y{z}^{4}x-4\,zx+3\,{y}^{2} \right) }}
\end{equation}
%
After loading the packages and `enter' the 2ODE (as in example 1), let's search a $S$-function. This time, typing
%
\begin{verbatim}
[> S1 := Sfunction(_2ode);
\end{verbatim}
results in an empty output. By trial and error we reach that for using {\tt Deg=9} we find:
\begin{verbatim}
[> S1 := Sfunction(_2ode,Deg=9);
\end{verbatim}
%
\begin{equation}
S1 :=  {\frac {x{z}^{7}+{z}^{3}y-1}{{z}^{2} \left( 3\,xy{z}^{4}-4\,xz+3\,{y}^{2} \right) }}
\end{equation}		
%
after $\approx 120 sec$ (and spending 320 MB of memory). However, if we use the parameter {\tt Sn=2} we can obtain the $S$-function $S_2$ in a much simpler way with
%
\begin{verbatim}
[> S2 := Sfunction(_2ode,Sn=2);
\end{verbatim}
%
\begin{equation}
S2 := {\frac {y}{x{z}^{2} \left( 3\,xy{z}^{4}-4\,xz+3\,{y}^{2} \right) }}
\end{equation}		
%
\begin{obs}
The degree is not increased automatically because in complicated cases a much higher degree can stop the computer due to lack of memory.
\end{obs}
%
From $S_2$ we can determine the associated 1ODEs. By typing
\begin{verbatim}
[> _1odeas := Exodes(_2ode,Sn=2,En=4);
\end{verbatim}
we get
\begin{eqnarray}
&&\_1odeas := {\frac {d}{dy}}z \left( y \right) =-{\frac {x z\left( y \right)^{7}+ z\left( y \right)^{3}y-1}{
 z\left( y \right)^{2} \left( 3\,xy z\left( y \right)^{4}-4\,xz \left( y \right) +3\,{y}^{2} \right) }}, \nonumber \\
&&{\frac {d}{dx}}z \left( x \right) =-{\frac {y}{x z\left( x \right)^{2} \left( 3\,xy z\left( x \right)^{4}-4\,xz \left( x \right) +3\,{y}^{2} \right) }}, \nonumber \\
&& {\frac {d}{dx}}y \left( x \right) =-{\frac {y \left( x \right) }{x \left( x{z}^{7}+{z}^{3}y \left( x \right) -1 \right) }} \nonumber
\end{eqnarray}		
%
We can try solving the associated 1ODEs with
\begin{verbatim}
[> H1 := Hfunction(_2ode,Sn=2);
[> H2 := Hfunction(_2ode,Sn=2,En=2);
[> H3 := Hfunction(_2ode,Sn=2,En=3);
\end{verbatim}
%
obtaining
\begin{eqnarray}
H1 & := & {z}^{3}y-\ln  \left( {z}^{4}x+y \right) \nonumber \\
H2 & := & -{\frac {{z}^{4}x+y}{x{{\rm e}^{{z}^{3}y}}}} \nonumber \\
H3 & := & -{\frac {x{{\rm e}^{{z}^{3}y}}}{{z}^{4}x+y}} \nonumber
\end{eqnarray}		
%
Constructing the operator $D_x$ with
\begin{verbatim}
[> DX := Dx(_2ode):
\end{verbatim}
%
and applying it to the $H$-functions
\begin{verbatim}
[> DX(H1); DX(H2); DX(H3);
\end{verbatim}
%
we have
\begin{eqnarray}
&& -\frac{1}{x} \nonumber \\ [2mm]
&&  \,\,\,0 \nonumber \\
&&  \,\,\,0 \nonumber
\end{eqnarray}		
%
We see that, in this case, we do not need to apply {\tt PDEassol} because the $H$-functions $H_2$ and $H_3$ directly give us the first integral we were looking for.
Again, we can try to completely solve the 2ODE by using the command
\begin{verbatim}
[> sol2ode := Gensol(_2ode,Sn=2);
\end{verbatim}
that leads to an empty output.
%
%

\section{Performance}
\label{performance}
\hspace\parindent

The theory behind the method developed here (to deal with rational 2ODEs presenting Liouvillian first integrals) is linked to the relationship between the integrating factors (which can be constructed with Darboux polynomials) and the symmetries of 2ODE. Interestingly, the method is very effective when dealing with a class of rational 2ODEs that are particularly difficult to be handled by the Lie and Darboux methods: rational 2ODEs presenting complicated Lie symmetries and / or presenting integrating factors made up by high degree Darboux polynomials. The analysis presented in this section is divided in three parts\footnote{In this paper all the computational data (time of running etc) was obtained on the same computer with the following configuration: Intel(R) Core(TM) i5-3337U @ 1.8 GHz.}:
\begin{itemize}
\item First, we will show a set of 2ODEs (just a sample of an entire class) to which the {\tt dsolve} (a Maple built-in command -- one of the most powerful ODE solvers) fails to solve / reduce.
\item In the second part of our analysis, we will make a comparison between our method and the Darbouxian and Lie approaches\footnote{Probably the most comprehensive methods for dealing with rational 2ODEs presenting Liouvillian first integrals.}.
\item Finally we will present some examples demonstrating the scope of the method and how we can use the package commands to facilitate the solution / study of more complicated 2ODEs
\end{itemize}

\subsection{A series of 2ODEs that are out of the scope of the {\tt dsolve} command}
\label{exten}
\hspace\parindent
In this section we present a series of 2ODEs where the powerful solver of the Maple (release 17) -- {\tt dsolve} command -- can not solve nor reduce. The Table \ref{tab_ds_x_m} below shows the CPU time and memory that the package {\it InSyDE} spends to find the first integrals (through the use of {\tt Invade} command).

\begin{table}[h]
{\begin{center} {\footnotesize
\begin{tabular}
{c|c|c|c|c|}
\hline
 &$\phi$ & $I$ & Time & Memory \\
\hline
 1 & $-{\frac {{x}^{2}yz-{x}^{2}{z}^{2}-x{y}^{3}-{y}^{2}zx-x{z}^{2}+{y}^{3}+
{y}^{2}z+2\,y{z}^{2}-{z}^{2}}{y \left( {x}^{2}-y \right) }}
$ & ${\frac { \left( -yx+z \right) {{\rm e}^{-x}}}{xz-{y}^{2}}}$ & 0.2 sec & $\approx$20MB   \\
\hline
 2 & $-{\frac {{x}^{2}{z}^{2}-2\,{y}^{2}x+2\,xyz-{z}^{2}}{{x}^{2}{y}^{2}-{x}
^{2}yz-{x}^{2}y-yz+{z}^{2}+y}}$ & ${\frac { \left( -{x}^{2}y+z \right) {{\rm e}^{-z}}}{-y+z}}$ & 0.3 sec & $\approx$20MB   \\
\hline
 3 & $-{\frac {xy{z}^{2}+x{z}^{3}+{y}^{2}z-y{z}^{2}-xz-y+2\,z}{{y}^{2}+x}}$ & ${\frac { \left( yz-1 \right)
{{\rm e}^{x}}}{zx+y}}$ & 0.05 sec & $\approx$20MB   \\
\hline
 4 & $-{\frac { \left( xyz+{y}^{2}-2\,y+2 \right) z}{ \left( y-1 \right)
 \left( zx-x+y \right) }}$ & ${\frac { \left( zx+y \right) {{\rm e}^{-z-y}}}{y-1}}$ & 0.3 sec & $\approx$20MB  \\
\hline
 5 & $-{\frac {{x}^{2}{z}^{3}+{z}^{2}xy+{x}^{2}z+yx-zx+y}{x \left( xyz+{y}^{
2}-x \right) }}$ & ${\frac { \left( zx+y \right) {{\rm e}^{-yz}}{{\rm e}^{-x}}}{x}}$ & 0.7 sec & $\approx$20MB  \\
\hline
 6 & $-{\frac { \left( zx+y-2\,z \right) z}{x{z}^{2}+yz+y}}$ & ${\frac { \left( zx+y \right) {{\rm e}^{-z-x}}}{z}}$ & 0.3 sec & $\approx$20MB  \\
\hline
 7 & $-{\frac { \left( xyz+{x}^{2}-zx+y \right) z}{x \left( yz+x-y \right) }
}$ & ${\frac { \left( yz+x \right) {{\rm e}^{-z-y}}}{x}}$ & 0.3 sec & $\approx$20MB  \\
\hline
 8 & $-{\frac {x{y}^{2}-{y}^{2}+yz+{z}^{2}}{y \left( yx+z-1 \right) }}$ & ${\frac { \left( yx+z \right) {{\rm e}^{-z}}{{\rm e}^{-x}}}{y}}$ & 0.2 sec & $\approx$20MB  \\
\hline
 9 & $-{\frac {{x}^{2}y-{x}^{2}z+x{y}^{2}+zx-{y}^{2}+yz+{z}^{2}+z}{ \left( y
+x \right)  \left( yx+z-1 \right) }}$ & ${\frac { \left( yx+z \right) {{\rm e}^{-z}}{{\rm e}^{-x}}}{y+x}}$ & 0.2 sec & $\approx$20MB  \\
\hline
10 & ${\frac {x{z}^{2}+{y}^{2}+yz-{z}^{2}}{x{y}^{2}+xyz+yx+yz+{z}^{2}-y}}$ & ${\frac { \left( yx+z \right) {{\rm e}^{-z}}}{z+y}}$ & 0.2 sec & $\approx$20MB  \\
\hline
\end{tabular} }
\end{center}}
\caption{In this table, we present 10 2ODEs of the form dz/dx=$\Phi$, a respective first integral and the time and memory consumed by our command in determinig this first integral.}
\label{tab_ds_x_m}
\end{table}

From observation of the Table \ref{tab_ds_x_m} we can see that the {\tt dsolve} command has difficulties when applied to rational 2ODEs presenting first integrals of the general form $e^{W_0}\prod W_i^{n_i}$, where the $W$s are polynomial functions. In the next section we will try to find reasons why our procedure is (for this class of rational 2ODEs) more efficient than the Darbouxian and Lie approaches.

\subsection{A comparison with the standard Darboux and Lie approaches}
\label{acsdla}
\hspace\parindent
Table \ref{tab_RsyS} shows the symmetries of the same 10 2ODEs $1-10$ presented on table \ref{tab_ds_x_m}. The results displayed there may provide a possible explanation for the low efficiency of the Lie method for this type of 2ODE: the symmetries are non local and very hard to obtain. On the other hand the integrating factors are formed with Darboux polynomials of degrees 2 and 3. The problematic cases for the Darbouxian approach are just the 2ODEs 2 and 5 (see Table \ref{tab_RsyS}) which present the Darboux polynomials of degree 3: 2ODE 2 - The Darboux approach could not find \footnote{using our package {\it FiOrDi}, available at the CPC program Library - http://cpc.cs.qub.ac.uk/, Catalogue identifier: AEQL-v1-0} the polynomial $x^2y-z$ in 300 sec (consuming 300MB); 2ODE 5 - The Darboux approach finds the polynomial $xyz+y^2-x$ in 120 sec (consuming 300MB).

\begin{table}[h]
{\begin{center} {\footnotesize
\begin{tabular}
{c|c|c|c|}
\hline
 & $R$ & $\nu$  & $S_1$ \\
\hline
 1 & ${\frac {{{\rm e}^{x}}y \left( {x}^{2}-y \right) }{ \left( yx-z
 \right) ^{2}}}$ &
${{\rm e}^{-\int \!-{\frac {{x}^{2}z+x{y}^{2}-2\,yz}{y \left( {x}^{2}-y
 \right) }}{dx}}}$ &
$-{\frac {{x}^{2}z+x{y}^{2}-2\,yz}{y \left( {x}^{2}-y \right) }}$
   \\
\hline
 2 & ${\frac {{x}^{2}{y}^{2}-{x}^{2}yz-{x}^{2}y-yz+{z}^{2}+y}{ \left( y-z
 \right)  \left( {x}^{2}y-z \right) }}$ &
${{\rm e}^{-\int \!{\frac {z \left( {x}^{2}-1 \right) }{{x}^{2}{y}^{2}-
{x}^{2}yz-{x}^{2}y-yz+{z}^{2}+y}}{dx}}}$ &
${\frac {z \left( {x}^{2}-1 \right) }{{x}^{2}{y}^{2}-{x}^{2}yz-{x}^{2}y
-yz+{z}^{2}+y}}$
  \\
\hline
 3 & $-{\frac {{{\rm e}^{x}} \left( {y}^{2}+x \right) }{ \left( xz+y
 \right) ^{2}}}$ &
${{\rm e}^{-\int \!{\frac {x{z}^{2}+1}{{y}^{2}+x}}{dx}}}$ &
${\frac {x{z}^{2}+1}{{y}^{2}+x}}$
  \\
\hline
 4 & ${\frac {{{\rm e}^{-z-y}} \left( xz-x+y \right) }{y-1}}$ &
${{\rm e}^{-\int \!{\frac {xyz+{y}^{2}-y+1}{ \left( y-1 \right)
 \left( xz-x+y \right) }}{dx}}}$ &
${\frac {xyz+{y}^{2}-y+1}{ \left( y-1 \right)  \left( xz-x+y \right) }}$
 \\
\hline
 5 & ${\frac {{{\rm e}^{-yz}}{{\rm e}^{-x}} \left( xyz+{y}^{2}-x \right) }{x
}}$ &
${{\rm e}^{-\int \!{\frac {x{z}^{2}+yz-1}{xyz+{y}^{2}-x}}{dx}}}$ &
${\frac {x{z}^{2}+yz-1}{xyz+{y}^{2}-x}}$
 \\
\hline
 6 & ${\frac {x{z}^{2}+yz+y}{z \left( xz+y \right) }}$ &
${{\rm e}^{-\int \!-{\frac {z}{x{z}^{2}+yz+y}}{dx}}}$ &
$-{\frac {z}{x{z}^{2}+yz+y}}$
 \\
\hline
 7 & ${\frac {{{\rm e}^{-z-y}} \left( yz+x-y \right) }{x}}$ &
${{\rm e}^{-\int \!{\frac {yz+x-z}{yz+x-y}}{dx}}}$ &
${\frac {yz+x-z}{yz+x-y}}$
 \\
\hline
 8 & ${\frac {{{\rm e}^{-z}}{{\rm e}^{-x}} \left( yx+z-1 \right) }{y}}$ &
${{\rm e}^{-\int \!{\frac {z}{y \left( yx+z-1 \right) }}{dx}}}$ &
${\frac {z}{y \left( yx+z-1 \right) }}$
 \\
\hline
 9 & ${\frac {{{\rm e}^{-z}}{{\rm e}^{-x}} \left( yx+z-1 \right) }{y+x}}$ &
${{\rm e}^{-\int \!-{\frac {{x}^{2}-z}{ \left( y+x \right)  \left( yx+z
-1 \right) }}{dx}}}$ &
$-{\frac {{x}^{2}-z}{ \left( y+x \right)  \left( yx+z-1 \right) }}$
 \\
\hline
10 & ${\frac {x{y}^{2}+xyz+yx+yz+{z}^{2}-y}{ \left( z+y \right)  \left( yx+z
 \right) }}$ &
${{\rm e}^{-\int \!-{\frac {z \left( x-1 \right) }{x{y}^{2}+xyz+yx+yz+{
z}^{2}-y}}{dx}}}$ &
$-{\frac {z \left( x-1 \right) }{x{y}^{2}+xyz+yx+yz+{z}^{2}-y}}$
 \\
\hline
\end{tabular} }
\end{center}}
\caption{In this table, we present, for the same 10ODEs presented on table (\ref{tab_ds_x_m}), the  Integrating Factors, Symmetries and $S$-functions}
\label{tab_RsyS}
\end{table}

The computation of Darboux polynomials of degrees higher than 3  is usually a computational `problem'. So, 2ODEs of this type (i.e., presenting first integrals that are Darboux functions or integrals of Darboux functions) can also create problems for the Darbouxian approach if the Darboux polynomials present on the integrating factor are of high degree (in practice greater than or equal to three). We can confirm this by observing the examples below: in Table \ref{sfif} we present the 2ODEs and the respective $S$-functions and integrating factors; then, in Table \ref{compDS}, we present a comparison between the time spent by the {\tt Sfunction} command for determining the $S$-function and the time for calculating the Darboux polynomials required for the construction of an integrating factor.

\begin{table}[h]
{\begin{center} {\footnotesize
\begin{tabular}
{c|c|c|c|}
\hline
 & $\phi$ & $S_1$ & $R$ \\
\hline
 11 & $-{\frac { \left( z{x}^{3}-3\,{x}^{2}z+y-z \right) z}{{x}^{3}{z}^{2}+zy+y}}$ & $-{\frac {z}{{x}^{3}{z}^{2}+zy+y}}$ & ${\frac {1}{z \left( z{x}^{3}+y \right)}}$  \\
\hline
 12 &  ${\frac {{x}^{5}z-{x}^{4}y-4\,{x}^{4}z+4\,{x}^{3}y+xy{z}^{2}-x{z}^{3}-{y}^{2}z+y{z}^{2}}{-({x}^{5}+{y}^{2})}}$ & $-{\frac {x \left( {x}^{3}+{z}^{2} \right) }{{x}^{5}+{y}^{2}}}
$ & ${\frac {1}{ \left( xz-y \right)  \left( {x}^{4}+zy \right) }}$  \\
\hline
 13 & $-{\frac {{x}^{2}{z}^{4}+xy{z}^{2}-xz+y}{x \left( xy{z}^{2}-2\,xz+{y}^{2} \right) }}$ & ${\frac {x{z}^{3}+zy-1}{xy{z}^{2}-2\,xz+{y}^{2}}}$ & ${\frac {1}{x \left( x{z}^{2}+y \right) }}$  \\
\hline
 14 & $-{\frac { \left( {x}^{4}-4\,{x}^{3}+y \right) z}{{x}^{4}+y}}$ & $-{\frac {z}{{x}^{4}+y}}$ & ${\frac {1}{z \left( {x}^{4}+y+z \right) }}$  \\
\hline
 15 & ${\frac {{z}^{5}{y}^{2}+y{z}^{5}+2\,{z}^{3}yx+x{z}^{3}-y{z}^{3}+{x}^{2}z+y{z}^{2}+x-y}{- 2\left( y{z}^{2}+x-y \right) zy}}$ & ${\frac {{y}^{2}{z}^{4}+y{z}^{4}+2\,xy{z}^{2}+x{z}^{2}-y{z}^{2}+{x}^{2}}{ 2\left( y{z}^{2}+x-y \right) zy}}$ & ${\frac {{{\rm e}^{ \left( y{z}^{2}+x \right) ^{-1}}}}{ \left( y{z}^{2}
+x \right) ^{2}}}$  \\
\hline
\end{tabular} }
\end{center}}
\caption{2ODEs, $S$-functions and Integrating Factors}
\label{sfif}
\end{table}

\begin{table}[h]
{\begin{center} {\footnotesize
\begin{tabular}
{c|c|c|c|c|c|c|}
\hline
 & \multicolumn{3}{|c|}{SM} & \multicolumn{3}{|c|}{DA} \\
\cline{2-7}
  & Time & Memory & Result & Time & Memory & Result  \\
\hline
 11 &  0.08 sec & $\approx$5MB & positive & 5 min & $\approx$300MB & negative  \\
\hline
 12 &  0.4 sec & $\approx$10MB & positive & 5 min & $\approx$300MB & negative  \\
\hline
 13 &  0.3 sec & $\approx$10MB & positive & 5 min & $\approx$300MB & negative  \\
\hline
 14 &  0.03 sec & $\approx$5MB & positive & 5 min & $\approx$300MB & negative  \\
\hline
 15 &  6.7 sec & $\approx$40MB & positive & 5 min & $\approx$300MB & negative  \\
\hline
\end{tabular} }
\end{center}}
\caption{$S$-function method $\times$ Darbouxian approach}
\label{compDS}
\end{table}

As we can see (and as expected), the 2ODEs 11 and 14 spent less time and memory (the degree of $P$ is 1) followed by the 2ODEs 12 and 13 (the degree of $P$ is 4). Finally, the 2ODE 15 was the most computationally costly (the degree of $P$ is 6). In order to expand the `range of action' and / or get a `better response' of the package, we can use a set of parameters. We will see this in the next section.

\subsection{The `special features' in action}
\label{sfa}
\hspace\parindent

The method used to find the polynomial $P$ (described briefly in section \ref{coam}) is called {\it method of undetermined coefficients} (MUC). This method tends to `explode' when the degree of the polynomials involved is high and when the resulting algebraic system is nonlinear. So, the package uses some extra-arguments to extend the scope of the procedure. Let's see them in action with an example:

Consider the rational 2ODE given by:
\begin{equation}
\label{perf32ode}
z' = -{\frac { \left( {x}^{5}y{z}^{2}+4\,{x}^{4}{y}^{2}z-x{z}^{2}+xz-4\,yz+
4\,y \right) z{x}^{3}}{{x}^{8}{y}^{2}{z}^{2}+{x}^{8}{y}^{2}z+zy{x}^{4}
+{x}^{4}y+1}}.
\end{equation}
After loading the package and the 2ODE (\ref{perf32ode}), we can try to find the first integral $I$ using the {\tt Invade} command. However, simply applying the command results in an empty output (even for high degrees of the polynomial $P$). By trial and error we reach that for using {\tt Deg=11} we can find $S_1$
\begin{verbatim}
[> S1 := Sfunction(_2ode,Deg=11);
\end{verbatim}
%
\begin{equation}
S1 := {\frac {z{x}^{4} \left( zy{x}^{4}-z+1 \right) }{{x}^{8}{y}^{2}{z}^{2}+{x}^{8}{y}^{2}z+zy{x}^{4}+{x}^{4}y+1}}
\end{equation}		
%
But only if we wait more than 400 seconds (and spending 500 MB of memory). This time interval is longer than we consider a failure to find the Darboux polynomials. However, if we use the parameter {\tt Den=x} and {\tt Sn=3} we can obtain the $S$-function $S_3$ in a much simpler way with
%
\begin{verbatim}
[> S3 := Sfunction(_2ode,Sn=3,Den=x);
\end{verbatim}
%
\begin{equation}
S3 := {\frac {4\,y}{x}}
\end{equation}		
%
in $0.07$ seconds. We can determine the associated 1ODEs by typing
\begin{verbatim}
[> _1odeas := Exodes(_2ode,Sn=3,En=4,Den=x);
\end{verbatim}
We obtain
\begin{eqnarray}
&&\_1odeas := {\frac {d}{dy}}z \left( y \right) =-{\frac {z\left( y \right) {x}^{4} \left( z\left( y \right) y{x}^{4}-z\left( y \right) +1 \right) }
{ z\left( y \right)^{2}{y}^{2}{x}^{8}+z\left( y \right) {y}^{2}{x}^{8}+z\left( y \right) y{x}^{4}+{x}^{4}y+1}}, \nonumber \\
&& {\frac {d}{dx}}z \left( x \right) ={\frac {-4\,z\left( x \right) {x}^{3}y \left( z\left( x \right) y{x}^{4}-z\left( x \right) +1 \right) }
{ z\left( x \right)^{2}{y}^{2}{x}^{8}+z\left( x \right) {y}^{2}{x}^{8}+z\left( x \right) y{x}^{4}+{x}^{4}y+1}}, \nonumber \\
&& {\frac {d}{dx}}y \left( x \right) ={\frac {-4\,y\left( x \right) }{x}} \nonumber
\end{eqnarray}		
%
We can try solving the associated 1ODE$_{[1]}$ with
\begin{verbatim}
[> H1 := Hfunction(_2ode,Sn=3,Den=x);
\end{verbatim}
that leads to an empty output. So, we can try 1ODE$_{[2]}$
\begin{verbatim}
[> H2 := Hfunction(_2ode,Sn=3,Den=x,En=2);
\end{verbatim}
%
that leads again to an empty output. Finally, we can try 1ODE$_{[3]}$
\begin{verbatim}
[> H3 := Hfunction(_2ode,Sn=3,Den=x,En=3);
\end{verbatim}
\begin{equation}
H3 := x^4y
\end{equation}		
Once we have found $H_3$, let's apply {\tt PDEassol}
\begin{verbatim}
[> pdeassol := PDEassol(_2ode,Sn=3,Den=x,En=3);
\end{verbatim}
%
\begin{eqnarray}
pdeassol &:=& \left[\frac{\partial H\left( z,h \right)}{\partial z} -{\frac {{h}^{2}{z}^{2}+{h}^{2}z+zh+h+1}{z \left( zh-z+1 \right) }}\,\frac{\partial H\left( z,h \right)}{\partial h}=0, \right. \nonumber \\
&& {\frac {d}{dz}}h \left( z \right) =-{\frac {{h(z)}^{2}{z}^{2}+{h(z)}^{2}z+zh(z)+h(z)+1}{z \left( zh(z)-z+1 \right) }}, \nonumber \\
&& \left. {\it \_H}_{{3}}={x}^{4}y,H\left( z,h \right) ={\rm e}^{\frac{1}{hz+1}}z+{\it Ei} \left( 1,- \frac{1}{hz+1}\right)  \right] \nonumber
\end{eqnarray}		
%
From the solution of the associated PDE$_{[3]}$, we can find a first integral $I$. Using the command
\begin{verbatim}
[>  Inv := Invade(_2ode,Sn=3,Den=x,En=3);
\end{verbatim}
we find
\begin{equation}
Inv := {\rm e}^{\frac{1}{x^4yz+1}}z+{\it Ei} \left( 1,- \frac{1}{x^4yz+1}\right)
\end{equation}
Again, we can try to completely solve the 2ODE by using the command
\begin{verbatim}
[> sol2ode := Gensol(_2ode,Sn=3,Den=x,En=3);
\end{verbatim}
%
that leads to an empty output.
%
%


\section{Conclusion}
\label{conclu}
\hspace\parindent
%
In \cite{Nosjpa2001}, some of us have introduced the so-called S-function, defined here on section (\ref{SandS}). It allowed us, at the time, to extend the Prelle-Singer procedure and produce a truly (semi)-algorithm to  deal with 2ODEs. It was a fruitful development and has generated many extensions and new developments either by us or by other researchers. Here, in this present paper, we have furthered the usage for this S-function.

Before embarking on emphasizing the practical advantages of the approach hereby introduced, we would like to comment a little further on the theoretical background we have used to reach these practical results.

In \ref{conectSandS}, we have dwelled on the concept of the associated 1ODEs. These have been introduced before in (\cite{Nosjmp2009}) but, in there, they served ``only'' as a theoretical auxiliary condition to prove a point. We have not noticed the practical value we have managed to assign to them here. Basically, here we have managed to use the fact that the Invariant for the 2ODE is a solution of the respective 1ODE (each one of these) to construct a way of doing the ``opposite'', {\it i.e.}, from the solutions to the 1ODEs find the invariant to the 2ODE. These results and methods were introduced in section (\ref{Iof2odes}).

 How come? The point is that, although the first order differential invariant for the 2ODE is a solution to the 1ODE, not all solutions of the 1ODE are a first order invariant for the 2ODE since (in turns) on of the variables ($x,y,z$) is regarded as a parameter, not a variable. But we can use the results presented on section (\ref{conectSandS}) and the PDE (\ref{calFeq}), introduced on the remark (\ref{solrelati}), to circumvent this situation and, from the general solution to the 1ODE, produce the solution to the 2ODE.

 All these results came about via the use of the ``$S$-function'' and its many properties hereby developed and explored. Here we have presented theoretical relation the $S$-function presents with the Lie symmetries and, in particular, equation (\ref{nuS}) encapsulates the explanation for the why our method hereby introduced is so much more efficient for some cases in finding the Lie symmetries. These are the cases of (some) non-local Lie symmetries. From equation (\ref{nuS}), once we have ``$S$'', we have the Lie symmetry. For the Darbouxian approach, the fact that our present method can be more efficient is clearer, it spurs from the fact that we managed to avoid the necessity of determinig the Darboux polynomials in order to determine the first order invariant.

So, somewhat ironically, the interplay the ``$S$-function'' presents with both approaches conspired to the fact that our method proved to be more efficient than both the Darbouxian and Lie methods in dealing with solving or reducing 2ODEs, for many cases.

 Even if some of our theoretical points were ``off'', the practical usage of the ideas and algorithms hereby introduced is demonstrated by the results in section (\ref{performance}), table (\ref{tab_ds_x_m}) and the two following ones on the section. These results validate the work here by themselves, but we believe in the theoretical conclusions drawn here and we are working on further developments of them,  mainly on the interplay of the ``$S$-function'' and the Lie symmetries.






\newpage
\begin{thebibliography}{25}


\bibitem{Lie}
S. Lie,
{\it Theorie der Transformationsgruppen},
{\bf Vol. I, II, III}, Chelsea, New York, (1970).

\bibitem{Dar}
G. Darboux,
{\it M\'emoire sur les \'equations diff\'erentielles alg\'ebriques du premier ordre et du premier
degr\'e (M\'elanges)},
Bull. Sci. Math. 2\`eme s\'erie 2, 60-96, 2, 123-144, 2, 151-200 (1878).

\bibitem{BluAnc}
G.W. Bluman and S.C. Anco,
{\it Symmetries and Integration Methods for Differential Equations},
Applied Mathematical Series {\bf Vol. 154}, Springer-Verlag, New York, (2002).

\bibitem{Ibr} N.H. Ibragimov,
{\it Elementary Lie Group Analysis and Ordinary Differential Equations},
Wiley: Chichester, (1999).

\bibitem{Olv}
P.J. Olver,
{\it Applications of Lie Groups to Differential Equations},
Springer-Verlag, (1986).

\bibitem{Sch}
F. Schwarz,
{\it Algorithmic Lie Theory for Solving Ordinary Differential Equations},
Chapman  Hall / CRC -- Taylor and Francis Group, (2008).

\bibitem{Ste}
W. H. Steeb,
{\it Continuous Symmetries, Lie Algebras, Differential Equations
and Computer Algebra},
World Scientific Publishing Co. Pte. Ltd., (2007)

\bibitem{PreSin}
M. Prelle and M. Singer,
{\it Elementary first integral ofdifferential equations.}
Trans. Amer. Math. Soc., {\bf  279} 215 (1983).

\bibitem{Noscpc1997} E.S. Cheb-Terrab, L.G.S. Duarte and L.A.C.P. da Mota,
{\it Computer Algebra Solving of First Order ODEs Using Symmetry Methods}.
Comput.Phys.Commun., {\bf 101}, 254, (1997).

\bibitem{Noscpc1998} E.S. Cheb-Terrab, L.G.S. Duarte and L.A.C.P. da Mota,
{\it Computer Algebra Solving of Second Order ODEs Using Symmetry Methods}.
Comput.Phys.Commun., {\bf 108}, 90, (1998).

\bibitem{AbrGuo}
B. Abraham-Shrauner and A. Guo,
{\it Hidden Symmetries Associated with the Projective
Group of Nonlinear First-Order Ordinary Differential Equations}.
J. Phys. A: Math.Gen., {\bf 25}, 5597-5608, (1992).

\bibitem{AbrGuo2}
B. Abraham-Shrauner and A. Guo,
{\it Hidden and Nonlocal Symmetries of Nonlinear Differential Equations},
Modern Group Analysis: Advanced Analytical and Computational Methods in
Mathematical Physics, Hidden Symmetries of Differential Equations
Editors: N.H. Ibragimov, M. Torrissi and G.A. Valenti, Dordrecht: Kluwer,
1-5, (1993).

\bibitem{AbrGovLea}
B. Abraham-Shrauner, K.S. Govinder and P.G.L Leach,
{\it Integration of second order ordinary differential equations not possessing
Lie point symmetries}.
Phys. Lett. A, {\bf 203}, 169-74, (1995).

\bibitem{Abr} B. Abraham-Shrauner, {\it Hidden symmetries and nonlocal
group generators for ordinary differential equations}.
IMA J. Appl. Math., {\bf 56}, 235-52, (1996).

\bibitem{GovLea} K.S. Govinder and P.G.L Leach,
{\it A group theoretic approach to a class of second-order ordinary differential
equations not possesing Lie point symmetries}.
J. Phys. A: Math. Gen., {\bf 30}, 2055-68, (1997).

\bibitem{AdaMah} A.A. Adam and F.M. Mahomed,
{\it Non-local symmetries of first-order equations}.
IMA J. Appl. Math., {\bf 60}, 187-98, (1998).

\bibitem{GanBru} M.L. Gandarias and M.S. Bruzón,
{\it Reductions for some ordinary differential equations through nonlocal symmetries}.
Journal of Nonlinear Mathematical Physics, Vol. 18, Suppl. 1, 123–133, (2011).

\bibitem{GanBruSen} M.S. Bruzón, M.L. Gandarias and M. Senthilvelan,
{\it Nonlocal symmetries of Riccati and Abel chains and their similarity reductions}.
 Journal of Mathematical Physics, {\bf 53}, 023512 (2012).

\bibitem{MurRom}
C. Muriel and J.L. Romero,
{\it New methods of reduction for ordinary differential equations},
IMA J. Appl. Math., {\bf 66}(2), 111-125, (2001).

\bibitem{MurRom2}
C. Muriel and J.L. Romero,
{\it $\,C^{\,\infty}$-Symmetries and reduction of equations without Lie
point symmetries},
J. Lie Theory, {\bf 13}(1), 167-188, (2003).

\bibitem{MurRom3}
C. Muriel and J.L. Romero,
{\it The $\lambda$-symmetry reduction method and Jacobi last multipliers},
Commun Nonlinear Sci Numer Simulat, {\bf 19} 807–820, (2014).

\bibitem{MurRom4}
C. Muriel and J.L. Romero,
{\it Nonlocal Symmetries, Telescopic Vector Fields and $\lambda$-Symmetries of Ordinary Differential Equations},
Symmetry, Integrability and Geometry: Methods and Applications, {\bf 8},  106-126, (2012).

\bibitem{CicGaeWal}
G. Cicogna, G. Gaeta and S. Walcher,
{\it Dynamical systems and $\sigma$-symmetries},
Journal of Physics A: Mathematical and Theoretical, {\bf 46}, Number 23 (2013).

\bibitem{CicGaeMor}
G. Cicogna, G. Gaeta and P. Morando,
{\it On the relation between standard and $\mu$-symmetries for PDEs},
Journal of Physics A: Mathematical and General, {\bf 37}, Number 40 (2004).

\bibitem{PucSac}
E. Pucci and G. Saccomandi,
{\it On the reduction methods for ordinary differential equations}.
J. Phys. A: Math. Gen., {\bf 35}, 6145-6155, (2002).

\bibitem{Nuc}
M.C. Nucci,
{\it Jacobi Last Multiplier and Lie Symmetries:
A Novel Application of an Old Relationship},
Journal of Nonlinear Mathematical Physics, {\bf 12}(2), 284-304, (2005).

\bibitem{Nuc2}
M.C. Nucci,
{\it Lie symmetries of a Painlevé-type equation without Lie symmetries},
Journal of Nonlinear Mathematical Physics, {\bf 15}(2), 205-211, (2008).

\bibitem{CaiLli}
L. Cairó and J. Llibre,
{\it Darboux Integrability for 3D Lotka-Volterra systems},
J. Phys. A: Math. Gen., {\bf 33} 2395-2406 (2000).

\bibitem{Chr} C. Christopher,
{\it Invariant algebraic curves and conditions for a center},
Proc. R. Soc. Edin. A, {\bf 124} 1209 (1994).

\bibitem{PreSin}
M. Prelle and M. Singer,
{\it Elementary first integral of differential equations.}
Trans. Amer. Math. Soc., {\bf  279} 215 (1983).

\bibitem{Sht}
R. Shtokhamer,
{\it Solving first order differential equations using the Prelle-Singer algorithm},
Technical report 88-09, Center for Mathematical Computation, University of Delaware (1988).

\bibitem{Col}
C. B. Collins,
{\it Algebraic Invariants Curves of Polynomial Vector Fields in the Plane,}
Preprint. Canada: University of Waterloo (1993); C B Collins,
{\it Quadratic Vector Fields Possessing a Centre},
Preprint. Canada: University of Waterloo (1993).

\bibitem{Sin}
M. Singer,
{\it Liouvillian First Integrals},
Trans. Amer. Math. Soc., {\bf  333} 673-688 (1992).

\bibitem{Chr2} C. Christopher,
{\it Liouvillian first integrals of second order polynomial differential equations}.
Electron. J. Differential Equations, No. 49, 7 pp. (electronic) (1999).

\bibitem{ChrLli}
C. Christopher and J. Llibre,
{\it Integrability via invariant algebraic curves for Planar polynomial differential systems},
Ann. Differential Equations, {\bf 16}, no. 1, 5-19 (2000).

\bibitem{Lli} J. Llibre,
{\it Integrability of polynomial differential systems, Handbook of Differential equations, Ordinary Differential Equations},
volume 1, Chapter 5, pages 437-531. Edited by A. Ca\~nada, P. Dr\'abek and A. Fonda. Elsevier B.V. (2004).

\bibitem{Nosjpa2002-1} L.G.S. Duarte, S.E.S.Duarte and L.A.C.P. da Mota,
{\it A method to tackle first order ordinary differential equations with Liouvillian functions in the solution},
J. Phys. A: Math. Gen., {\bf 35}, 3899-3910, (2002).

\bibitem{Nosjpa2002-2} L.G.S. Duarte, S.E.S.Duarte and L.A.C.P. da Mota,
{\it Analyzing the Structure of the Integrating Factors for First Order Ordinary Differential Equations with Liouvillian Functions in the Solution},
J. Phys. A: Math. Gen., {\bf 35}, 1001-1006, (2002).

\bibitem{Noscpc2002}  L.G.S. Duarte, S.E.S.Duarte, L.A.C.P. da Mota and J.F.E. Skea,
{\it Extension of the Prelle-Singer Method and a MAPLE implementation},
Computer Physics Communications, {\bf 144}, n. 1, 46-62, (2002).

\bibitem{Nosjcam2005} J. Avellar, L.G.S. Duarte, S.E.S. Duarte, L.A.C.P. da Mota,
{\it Integrating First-Order Differential Equations with Liouvillian Solutions via Quadratures: a Semi-Algorithmic Method,}
Journal of Computational and Applied Mathematics {\bf 182}, 327-332, (2005).

\bibitem{Nosjpa2001}
L.G.S. Duarte, S.E.S.Duarte, L.A.C.P. da Mota and J.F.E. Skea,
{\it Solving second order ordinary differential equations by extending the Prelle-Singer method},
J. Phys. A: Math.Gen., {\bf 34}, 3015-3024, (2001).

\bibitem{Noscpc2007}
J. Avellar, L.G.S. Duarte, S.E.S.Duarte and L.A.C.P. da Mota,
{\it  Determining Liouvillian first integrals for dynamical systems in the plane},
Computer Physics Communications, {\bf 177}, 584-596, (2007).

\bibitem{Nosamc2007}
J. AvellarL.G.S. Duarte, S.E.S.Duarte and L.A.C.P. da Mota,
{\it A semi-algorithm to find elementary first order invariants of rational second order ordinary differential equations},
Appl. Math. Comp., {\bf 184} 2-11 (2007).

\bibitem{Nosjmp2009}
L.G.S.Duarte and L.A.C.P.da Mota,
{\it Finding Elementary First Integrals for Rational Second Order Ordinary Differential Equations},
J. Math. Phys., {\bf 50}, (2009).

\bibitem{Nosjpa2010}
L.G.S.Duarte and L.A.C.P.da Mota,
{\it Finding Elementary First Integrals for Rational Second Order Ordinary Differential Equations},
J. Phys. A: Math. Theor. {\bf 43}, n.6, (2010).

\bibitem{LliZha}
J. Llibre and X. Zhang,
{\it Darboux theory of integrability for polynomial vector fields in $R^n$ taking into account the multiplicity at infinity},
Bull. Sci. Math. {\bf 133}, 765–778, (2009).

\end{thebibliography}


\end{document}